\documentclass[aps,twoside,superscriptaddress,showkeys,longbibliography,nofootinbib]{revtex4-2}

\usepackage{amsmath,amsfonts,amssymb,amsthm,mathtools}

\usepackage[english]{babel}
\usepackage[utf8]{inputenc}
\usepackage[T1]{fontenc}

\usepackage[dvipsnames]{xcolor}

\usepackage{subfigure}

\usepackage{graphicx}
\usepackage{siunitx}
\usepackage{comment}

\usepackage{enumitem}

\usepackage[
  unicode=true, 
  bookmarks=false,
  breaklinks=false,
  pdfborder={0 0 1},
  colorlinks=false]{hyperref}

\hypersetup{
  pdftitle={},
	pdfauthor={},
	colorlinks=true,
	linkcolor=blue,
	citecolor=red,
	filecolor=black,
	urlcolor=blue,
	bookmarksopen,
	bookmarksopenlevel=1,
}
\usepackage{hyperref}

\def\ii{\mathrm{i}}
\def\ee{\mathrm{e}}

\def\sgn{\operatorname{sgn}}
\newcommand{\pdag}{^{\vphantom{\dagger}}}
\newcommand{\ppr}{^{\vphantom{\prime}}}

\renewcommand{\Re}{\operatorname{Re}}

\def\v{v}
\def\vnull{v_{0}}
\def\pp{\hat{\varphi}}
\def\dd{\delta\hat{\rho}}
\def\nGP{n_\mathrm{GP}}

\def\QN{\hat{N}}
\def\KF{\hat{U}}
\def\pif{\hat{\pi}}

\newcommand{\wick}[1]{\left. :\! \hspace{-0.5pt} #1 \hspace{-0.5pt} \!: \right.}
 
\begin{document}


\title{%
Breaking of Huygens-Fresnel principle in inhomogeneous Tomonaga-Luttinger liquids%
}%

\author{Marek Gluza}
\email[]{marekludwik.gluza@ntu.edu.sg}
\affiliation{%
Dahlem Center for Complex Quantum Systems,
Freie Universit{\"a}t Berlin,
Arnimallee 14,
14195 Berlin, Germany%
}%
\affiliation{%
School of Physical and Mathematical Sciences,
Nanyang Technological University,
21 Nanyang Link,
637371 Singapore, Republic of Singapore%
}%

\author{Per Moosavi}
\email[]{pmoosavi@phys.ethz.ch}
\affiliation{%
Institute for Theoretical Physics,
ETH Zurich,
Wolfgang-Pauli-Strasse 27,
8093 Z{\"u}rich, Switzerland%
}%

\author{Spyros Sotiriadis}
\email[]{spyros.sotiriadis@fu-berlin.de}
\affiliation{%
Dahlem Center for Complex Quantum Systems,
Freie Universit{\"a}t Berlin,
Arnimallee 14,
14195 Berlin, Germany%
}%
\affiliation{%
Department of Physics,
Faculty of Mathematics and Physics,
University of Ljubljana,
Jadranska 19,
1000 Ljubljana, Slovenia%
}%

\date{%
January 20, 2022%
}%


\begin{abstract}
Tomonaga-Luttinger liquids (TLLs) can be used to effectively describe one-dimensional quantum many-body systems such as ultracold atoms, charges in nanowires, superconducting circuits, and gapless spin chains. Their properties are given by two parameters, the propagation velocity and the Luttinger parameter. Here we study inhomogeneous TLLs where these are promoted to functions of position and demonstrate that they profoundly affect the dynamics: In general, besides curving the light cone, we show that propagation is no longer ballistically localized to the light-cone trajectories, different from standard homogeneous TLLs. Specifically, if the Luttinger parameter depends on position, the dynamics features pronounced spreading into the light cone, which cannot be understood via a simple superposition of waves as in the Huygens-Fresnel principle. This is the case for ultracold atoms in a parabolic trap, which serves as our main motivation, and we discuss possible experimental observations in such systems.
\end{abstract}



\maketitle


{%
\hypersetup{hidelinks}
\tableofcontents
}%


\section{Introduction}


Conformal field theory (CFT) in one spatial dimension can successfully be used to effectively describe the low-energy regime of a large class of gapless systems commonly referred to as Tomonaga-Luttinger liquids (TLLs) \cite{Tomonaga1950, Luttinger1963, MattisLieb1965, Haldane1981, Haldane1981b}.
Examples of such systems include ultracold atoms, charges in nanowires, superconducting circuits, and certain spin chains, and their properties within the standard homogeneous TLL description can be encoded in two positive parameters, namely the propagation velocity $v$ and the so-called Luttinger parameter $K$, see \cite{Giamarchi2003, CCGOR2011} for a review.

\begin{figure}[!htbp]

\centering
    
\includegraphics[width=0.7\linewidth]{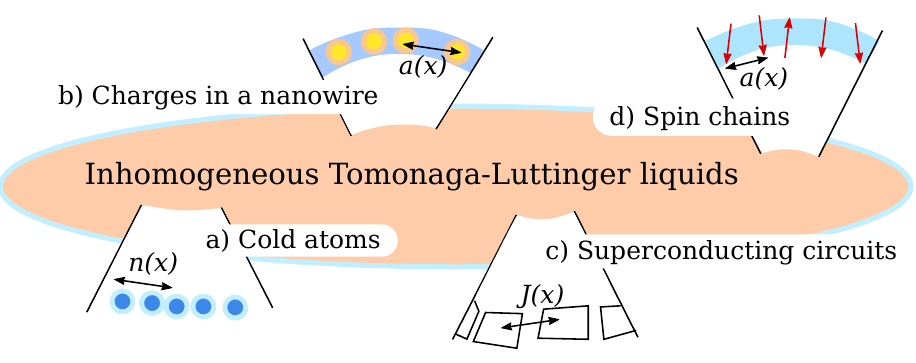}
    
\caption{%
Examples of applications of the inhomogeneous TLL description:
a) Ultracold atoms with mean density $n(x)$ varying with position $x$ (this is discussed in detail in Sec.~\ref{Sec:Atom_chip}).
b) Bent nanowire, where the bending leads to a spatially inhomogeneous lattice spacing $a(x)$.
c) Superconducting circuits with a spatially inhomogeneous coupling $J(x)$ between adjacent sites, assuming a spatially homogeneous coupling leads to a corresponding homogeneous TLL description.
d) Spin chains of the TLL universality class on a bent substrate crystal, again leading to a spatially inhomogeneous lattice spacing $a(x)$.%
}%

\label{Fig:Illustration}

\end{figure}

Here we extend this description to inhomogeneous systems where $v$ and $K$ are promoted to functions $v(x)$ and $K(x)$ of the spatial coordinate $x$.
Such a generalization appeared earlier in \cite{MaslovStone1995, SafiSchulz1995, Ponomarenko1995, GutmanGefenMirlin2010} to model quantum wires coupled to leads, and systems of this type have recently garnered considerable interest, in particular when studied out of equilibrium, see, e.g., \cite{ADSV2016, Dubail2017tonks, Dubail2017lightcones, BrunDubail2018igff, GLM2018, LangmannMoosavi2019, Moosavi2021, Ruggiero2019breathing, Murciano2019, Bastianello2020ill, RCDD2020}.
In Fig.~\ref{Fig:Illustration} we illustrate a selection of experimental setups that lend themselves to a TLL description (when homogeneous on a microscopic scale) and give possible sources of inhomogeneities that would lead to an \emph{inhomogeneous TLL} description, some of which have been studied in various detail before, see, e.g., \cite{Whitlock2003, MoraCastin2003, Cazalilla2004, GritsevPRB2007, Geiger2014, Dubail2017tonks, Dubail2017lightcones, LangmannMoosavi2019, Moosavi2021}.
We also mention \cite{KaneFischer1992PRL, KaneFischer1992PRB, GiamarchiSchulz1988} for related but different realizations of inhomogeneities in one-dimensional systems.

The most general Hamiltonian for inhomogeneous TLLs is of the form (we use units where $\hbar = 1$)
\begin{equation}
\label{Eq:H_iTLL}
H_{\mathrm{iTLL}}
= \frac{1}{2} \int_{-R}^{R} \mathrm{d}x\,
    \! \wick{
	    \biggl(
		    \frac{v(x)}{K(x)} \pif(x)^2
		    + v(x)K(x) [\partial_{x} \hat{\phi}(x)]^2
	    \biggr)
	} \!,
\end{equation}
where $2R$ is the system length, $\wick{\cdots}$ denotes (boson) normal ordering,
$[\hat{\phi}(x), \pif(x')] = \ii \delta(x - x')$,
and
$[\hat{\phi}(x), \hat{\phi}(x')]
= 0
= [\pif(x), \pif(x')]$.
Here, $\pif(x)$ and $\partial_{x} \hat{\phi}(x)$ have interpretations as the linearized particle density and current, respectively, within the harmonic-fluid (or phenomenological-bosonization) approach \cite{Haldane1981, Cazalilla2004, Giamarchi2003}.
We will assume that $R$ is finite throughout this paper:
Boundary conditions are addressed in Sec.~\ref{Sec:SL_theory} and left unspecified for now.
In Sec.~\ref{Sec:SL_theory} we also make precise in Fourier space that the above normal-ordering prescription means that creation operators are to the left of annihilation operators in any given term.

The Hamiltonian in \eqref{Eq:H_iTLL} can be used to describe a wide range of inhomogeneous settings using $v(x)$ and $K(x)$ that vary smoothly in space, see Fig.~\ref{Fig:Illustration} and references above, and even quantum wires coupled to leads if they are allowed to vary non-smoothly.
The latter case will not be studied here, as we will focus on the use of \eqref{Eq:H_iTLL} to model inhomogeneous TLLs.

One special family of inhomogeneous TLLs consists of those where only $v(x)$ depends on $x$ while $K$ is constant: In this case, all terms in the Hamiltonian are deformed by the same function $v(x)$ compared to the homogeneous case.
An example is a quantum spin chain of the usual TLL class (in the homogeneous case) where all couplings are generalized to vary smoothly in space in the same way, see, e.g., \cite{Dubail2017tonks, Dubail2017lightcones, LangmannMoosavi2019, Moosavi2021}.
The resulting effective description is a CFT with a position-dependent propagation velocity $v(x)$, referred to as inhomogeneous CFT, and has been studied both by curved-spacetime approaches \cite{Dubail2017tonks, Dubail2017lightcones, Ruggiero2019breathing, Murciano2019} and representation-theoretic tools \cite{GLM2018, LangmannMoosavi2019, Moosavi2021, LapierreMoosavi2021}.
One manifestation of this $v(x)$ is that excitations propagate along curved light-cone trajectories.
However, an $x$-dependent $K(x)$ is sometimes an inescapable consequence of the effective description of a given inhomogeneous system, see, e.g., \cite{BrunDubail2018igff, ADSV2016, Bastianello2020ill}, and is also known to make the model more difficult to solve: One way to see this is that such a $K(x)$ couples the right- and left-moving sectors of the CFT corresponding to a constant $K$, meaning that the resulting theory is no longer a CFT.

In this paper we consider another special family of inhomogeneous TLLs where
\begin{equation}
\label{vK_condition}
v(x)/K(x) = \v/K
\end{equation}
for some constants $\v$ and $K$.
Our main motivation comes from experiments on trapped ultracold atoms on an atom chip (see Sec.~\ref{Sec:Atom_chip} for details) using a strongly anisotropic trap such that the system is effectively one-dimensional.
In this example, the $x$-dependence of both $v(x)$ and $K(x)$ is dictated by the mean atom-density distribution, and the effective Hamiltonian is precisely of the form in \eqref{Eq:H_iTLL}, see, e.g., \cite{Schweigler2017, Rauer2018}.
The condition in \eqref{vK_condition} is natural since changing the mean atom-density distribution affects $v(x)$ and $K(x)$ in the same way, as it would be for any other inhomogeneous TLL where these functions are given by the same underlying position-dependent quantity.

While equilibrium results are interesting in their own right and have been studied before \cite{Stringari1996, HoMa1999, MenottiStringari2002, Ghosh2004, Petrov2004, Citro2008}, our primary focus is the effect of $K(x)$ on the non-equilibrium dynamics:
We will show that an $x$-dependent $K(x)$ implies that an initial localized wave packet does not propagate ballistically localized to the light cone but instead also spreads inside the cone.
When this is the case the Huygens-Fresnel principle is not valid, while we will show that it remains valid if only $v(x)$ depends on position.

We recall that the Huygens-Fresnel principle says that every point on a wavefront serves as the source of secondary waves and that a wavefront at a later time is the sum of the secondary waves emanating from the wavefront at an earlier time.
The principle can be expressed in mathematical terms as follows:
A linear hyperbolic partial differential equation satisfies the Huygens-Fresnel principle if its Green's function vanishes everywhere except on the hypersurface of the wavefront, see, e.g., \cite{Hadamard1923, Veselov1998}.
In physical terms, this means that a signal generated by an instantaneous disturbance will remain instantaneous throughout its transmission.
We also recall that this principle is valid only when the number of spatial dimensions is odd \cite{Hadamard1923}; we refer to \cite{Veselov1998} for an overview of the history.
The Huygens-Fresnel principle can be seen to be a characteristic feature of standard CFT dynamics in $1+1$ dimensions, as the time evolution of the fields $\partial_{x} \hat{\phi}$ and $\pif$ is expressed in terms of right- and left-moving fields that propagate strictly along the light cone.
We also mention that there are studies of violation of the Huygens-Fresnel principle in higher spatial dimensions in the context of cosmology and quantum information theory, see, e.g., \cite{Sonego1992, Faraoni1992, Blasco2015, Jonsson2015}.

\begin{figure}[!htbp]

\centering

\subfigure[Inhomogeneous TLL]{%
\includegraphics[height=0.4\linewidth]{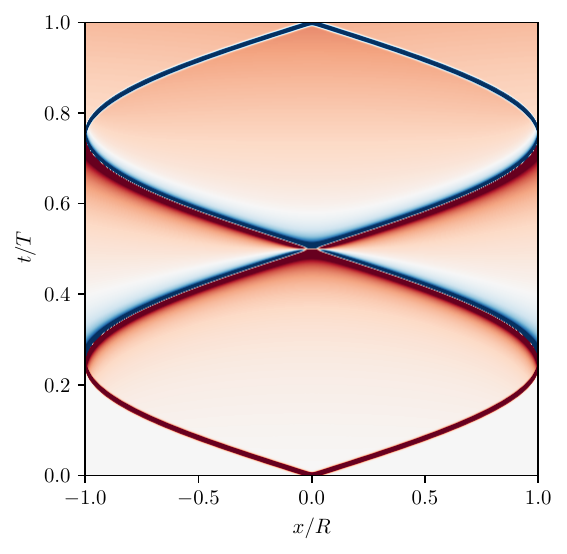}
\label{Fig:Propagation_of_Gaussian_wp_introduction:iTLL}
}%
\hspace{10mm}
\subfigure[Inhomogeneous-velocity wave equation]{%
\includegraphics[height=0.4\linewidth]{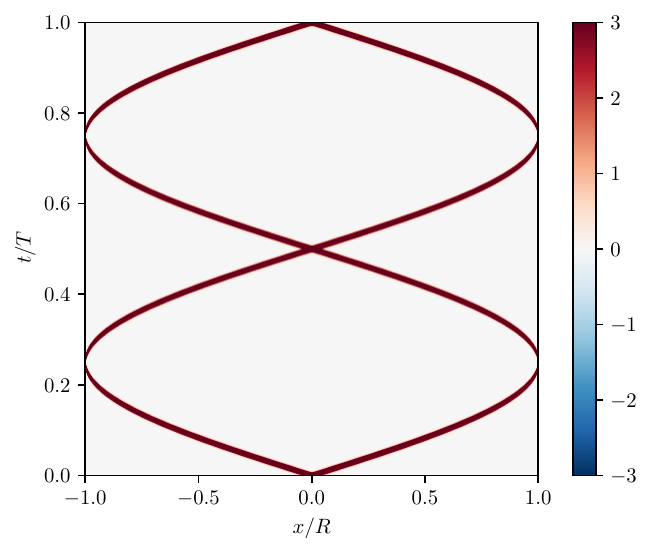}
\label{Fig:Propagation_of_Gaussian_wp_introduction:iVWE}
}%

\caption{%
Propagation of an initial localized Gaussian wave packet for
\subref{Fig:Propagation_of_Gaussian_wp_introduction:iTLL} inhomogeneous TLL with $v(x)$ and $K(x)$ given by \eqref{K_parabolic} (see Sec.~\ref{Sec:Dynamics_for_iTLLs:GF} for details)
and
\subref{Fig:Propagation_of_Gaussian_wp_introduction:iVWE} inhomogeneous-velocity wave equation (corresponding to inhomogeneous CFT mentioned in the text) with $v(x)$ as in \eqref{K_parabolic} but constant Luttinger parameter $K(x) = K$ (see Sec.~\ref{Sec:Comparison_iVWE} for details).
In \subref{Fig:Propagation_of_Gaussian_wp_introduction:iTLL} the wave packet splits into two counter-propagating packets that follow curved light-cone trajectories with non-zero contribution inside the cone, indicating the presence of long tails and the breaking of the Huygens-Fresnel principle, while in \subref{Fig:Propagation_of_Gaussian_wp_introduction:iVWE} the packets are ballistically localized to the edges of the cone.
Note that the former exhibits a sign inversion after one (quasi-)period $T = 2\pi R/v$ while the latter does not.
}%

\label{Fig:Propagation_of_Gaussian_wp_introduction}

\end{figure}

In the present case of inhomogeneous TLLs with an $x$-dependent Luttinger parameter $K(x)$, we argue that the spreading in the interior of the light cone is universal (i.e., independent of microscopic details), depending only on that the effective description has such a $K(x)$.
In contrast, the effect of $v(x)$ is to deform the light cone itself. 
As a specific example, we consider the special case
\begin{equation}
\label{K_parabolic}
K(x)/K = v(x)/\v = \sqrt{1 - (x/R)^2}.
\end{equation}
This arises, e.g., in the weakly interacting regime of an ultracold atomic gas confined by a parabolic trapping potential, which is one of the most common settings in cold-atom experiments (again, see Sec.~\ref{Sec:Atom_chip} for details).
As we will see, in this case it is natural to expand the fields in Legendre polynomials instead of cosine waves (assuming Neumann boundary conditions) in order to diagonalize the Hamiltonian, see, e.g., \cite{Petrov2004}.
This in turn will be used to compute analytical results for correlation functions.
Alternatively, by deriving and solving the equations of motion for the fields, one can directly visualize the spreading inside the light cone and thus the breaking of the Huygens-Fresnel principle, as shown in Fig.~\ref{Fig:Propagation_of_Gaussian_wp_introduction}.
To our knowledge, the possibility to violate the Huygens-Fresnel principle in inhomogeneous TLLs is new and deserves receiving attention due to the experimental importance of TLLs.

The method that we will present to diagonalize the inhomogeneous TLL Hamiltonian is based on Sturm-Liouville theory.
This naturally applies to more general $v(x)$ and $K(x)$ than those in \eqref{K_parabolic}, and also includes the homogeneous case, as will become clear.
However, we choose the parabolic case in \eqref{K_parabolic} as our main example since it directly corresponds to a common trapping potential for ultracold atoms and since it leads to eigenmode expansions using Legendre polynomials, while other possible examples would lead to expansions in other sets of orthonormal eigenfunctions.

As a last remark, it is instructive to note that in the path-integral formulation, the Hamiltonian in \eqref{Eq:H_iTLL} is equivalent to the Lagrangian density
\begin{equation}
\label{cL_iTLL}
\mathcal{L}
= \frac{\v}{2} \sqrt{-h} K(x)
  h^{\mu\nu} (\partial_{\mu} \phi) (\partial_{\nu} \phi)
\end{equation}
in curved spacetime described by the metric
$(h_{\mu\nu}) = \operatorname{diag}(v(x)^2/\v^2, -1)$
using the coordinates $(x^0, x^1) = (\v t, x)$ (see Appendix~\ref{App:Lagrangian_in_curved_spacetime} for details).
As can be seen, $K(x)$ appears in $\mathcal{L}$ directly and not in the metric.
Indeed, it cannot be absorbed in $(h_{\mu\nu})$, indicating that its effect is different from $v(x)$, which is purely geometric from this point of view.

The remainder of this paper is organized as follows.
In Sec.~\ref{Sec:Atom_chip} we further elaborate on ultracold atoms trapped on an atom chip (which serves as our main experimental motivation).
In Sec.~\ref{Sec:SL_theory} we present the method for diagonalizing the inhomogeneous TLL Hamiltonian based on Sturm-Liouville theory.
This is used in Sec.~\ref{Sec:Dynamics_for_iTLLs} to compute non-equal-time correlation functions and to study the Heisenberg equations of motion as well as compute and analyze their corresponding Green's functions.
In Sec.~\ref{Sec:Comparison_iVWE} we compare the results for inhomogeneous TLLs with those for the corresponding case where only $v(x)$ depends on position.
In Sec.~\ref{Sec:Experimental_observations} we discuss the possibility of experimentally observing violation of the Huygens-Fresnel principle in ultracold atoms.
Concluding remarks are given in Sec.~\ref{Sec:Concluding_remarks}.


\section{Ultracold atoms on an atom chip}
\label{Sec:Atom_chip}


The field theoretic model considered in this paper can be quantum simulated using a one-dimensional quasi-condensate of ultracold atoms confined in an external trap \cite{Whitlock2003, MoraCastin2003, Cazalilla2004, GritsevPRB2007, Geiger2014}.
This has been achieved in a highly accurate and versatile manner using an atom chip (an electronic device that allows for precise control of the external trap) that can be used to tune the parameters of the effective field theory model over a wide range of values \cite{Langen2015, Schweigler2017, Zache2020, Schweigler2020decay}. 
In addition, using the recently added digital micromirror device (DMD) functionality, it is now possible to engineer space- and time-dependent external potentials of arbitrary shape, which translate into an inhomogeneous mean atom-density distribution \cite{Tajik2019}.

More specifically, a one-dimensional gas of ultracold atoms interacting through a short-range inter-particle potential can be effectively described at low energies using the harmonic-fluid approach, which is given by the following Hamiltonian of standard TLL form (we omit normal ordering in this section for simplicity):
\begin{equation}
\label{Eq:H0_atoms}
H_{\mathrm{G}}
= \frac{1}{2}
  \int_{-R}^{R} \mathrm{d}x\,
  \biggl(
    g [\dd(x)]^2
    + \frac{\nGP}{m} [\partial_{x} \pp(x)]^2
  \biggr).
\end{equation}
Here, the field $\pp(x)$ corresponds to the phase of the condensate and the field $\dd(x)$ to the deviation of the density away from the mean value $\nGP$, where $x$ is the position in the longitudinal direction. 
These fields are bosonic and canonically conjugate in the sense that $[\dd(x), \pp(x')] = \ii \delta(x - x')$
and
$[\dd(x), \dd(x')] = 0 = [\pp(x), \pp(x')]$.
The parameter $m$ denotes the atomic mass, and $g$ is proportional to the scattering length of the atoms and also depends on other characteristics of the trap, specific to the experimental implementation \cite{Rauer2018}.

In the presence of a non-uniform external trap, the mean atom density becomes space dependent, and its distribution $\nGP(x)$ can be obtained from the Gross-Pitaevskii (GP) equation, see, e.g., \cite{MoraCastin2003}.
As mentioned, on the atom chip, the shape of $\nGP(x)$ can be programmably controlled using a DMD \cite{Tajik2019}.
Assuming that the trap potential varies slowly compared to the length scale of the mean inter-particle distance, the low-energy physics of the system is still described by the above Hamiltonian with $\nGP$ replaced by $\nGP(x)$, which we will see is precisely an example of the inhomogeneous TLL model in \eqref{Eq:H_iTLL} [cf.\ \eqref{Eq:Conversion}--\eqref{v_x_K_x_nGP_x} below].

Moreover, a pair of parallel and adjacent one-dimensional quasi-condensates separated by a potential barrier in the transverse direction can be used to realize the sine-Gordon model \cite{Gritsev2006, GritsevPRB2007, Schweigler2017, Kukuljan2018correlation}.
In this case the coupling between the two condensates (due to atomic hopping through the barrier) plays the role of a Josephson junction corresponding to a cosine self-interaction of the relative phase field, namely the field describing the phase difference $\pp_{\mathrm{r}}(x)$ between the two condensates.
The relative part of the effective field-theory description is given by the Hamiltonian
\begin{equation}
\label{Eq:H_sG}
H_{\mathrm{sG}}
= \int_{-R}^{R} \mathrm{d}x\,
  \biggl(
    g [\dd_{\mathrm{r}}(x)]^2
    + \frac{\nGP(x)}{4m} [\partial_{x} \pp_{\mathrm{r}}(x)]^2
    - 2 J \nGP(x) \cos(\pp_{\mathrm{r}}(x))
  \biggr),
\end{equation}
where $\dd_{\mathrm{r}}(x)$ is the relative density deviation, corresponding to the difference between the density deviations of the two condensates,
and is conjugate to $\pp_{\mathrm{r}}(x)$.
In the above, the parameter $J$ is the strength of the tunnel coupling between the two condensates, which can be tuned by adjusting the barrier of the transversal double-well trapping potential; see, e.g., \cite{Rauer2018, Schweigler_thesis} for a detailed discussion in relation to a recent experiment or the theoretical modeling in \cite{Whitlock2003, GritsevPRB2007}.
Note that the full description includes not only the relative fields but also the common fields corresponding to the sums of the density deviations and phases, respectively, of the two condensates, including possible higher-order couplings between the relative and common fields \cite{GritsevPRB2007}.
However, it is possible to motivate studying only the relative part as the effects of such couplings have been observed to be rather weak \cite{Rauer2018, vanNieuwkerk2020}.

As observed in atom-chip experiments, cooling the gas in the presence of intermediate values of tunnel coupling between the two adjacent quasi-condensates leads to a preparation of non-Gaussian states while for small and large values the states are effectively Gaussian.
In the uncoupled case, $J=0$, the system is described by the Gaussian TLL model [though with different parameter values compared to \eqref{Eq:H0_atoms}, see below].
On the other hand, in the strongly coupled case, $J\to\infty$, it is described by the effective Gaussian model obtained by replacing $\cos(\pp_{\mathrm{r}}(x))$ in \eqref{Eq:H_sG} by $1 - \pp_{\mathrm{r}}(x)^2/2$ (and renormalizing away the constant term), see, e.g., \cite{Rauer2018}.
We remark that higher-order terms of the expansion of this cosine term can be taken into account by truncation methods, see, e.g., \cite{FRT1998, FRT1999, Kukuljan2018correlation, JamesEtAl2018}.

Thus, both for one quasi-condensate and for a pair of quasi-condensates in the above two limits, defining
\begin{equation}
\label{Eq:Conversion}
\hat{\phi}(x)
= - \sqrt{2/\pi} \pp(x)
= - \sqrt{1/\pi} \pp_{\mathrm{r}}(x),
\quad
\pif(x)
= \sqrt{\pi/2} \dd(x)
= \sqrt{\pi} \dd_{\mathrm{r}}(x),
\quad
\v/K
= 2g/\pi,
\quad
M\v
= 2\sqrt{Jm},
\end{equation}
where $M$ is an effective mass, we obtain a Hamiltonian of the form
\begin{equation}
H
= H_{\mathrm{iTLL}}
  + \frac{(M\v)^2}{2} \int_{-R}^{R} \mathrm{d}x\,
    v(x) K(x) \hat{\phi}(x)^2
\end{equation}
with $H_{\mathrm{iTLL}}$ in \eqref{Eq:H_iTLL} and
\begin{equation}
\label{v_x_K_x_nGP_x}
v(x)
= \sqrt{\frac{\nGP(x) g}{m}},
\qquad
K(x)
= \frac{\pi}{2} \sqrt{\frac{\nGP(x)}{mg}}.
\end{equation}
Note that the latter satisfy the condition in \eqref{vK_condition} with $v/K$ in \eqref{Eq:Conversion}.

The above discussion provides an intuitive dictionary to understand which parameters can be most readily engineered in atom-chip experiments.
In principle, from this point of view, $\nGP(x)$ is obtained by solving a non-linear equation.
It is, however, instructive to consider the Thomas-Fermi (TF) approximation, see, e.g., \cite{Dunjko2001}, which is valid in the weakly interacting regime and predicts that
\begin{equation}
\nGP(x) \approx \frac{\mu - V(x)}{g},
\label{Eq:TF}
\end{equation}
where $\mu$ is the chemical potential fixed by the average number of atoms in the system and $V(x)$ is the trapping potential.
For a parabolic trapping potential the approximate mean atom-density distribution is thus given by
\begin{equation}
\nGP(x)
\approx \nGP^\mathrm{max} \bigl[ 1 -  (x/R)^{2} \bigr],
\end{equation}
where $R$ is the TF radius fixed by the peak density $\nGP^\mathrm{max}$ and the parabolic trapping frequency.
This gives precisely our special case in \eqref{K_parabolic} if we set $K = (\pi/2) \sqrt{\nGP^\mathrm{max}/mg}$.


\section{Diagonalization of the inhomogeneous TLL Hamiltonian}
\label{Sec:SL_theory}


In this section we diagonalize the inhomogeneous TLL Hamiltonian by expanding the fields using eigenmodes obtained by solving the associated Sturm-Liouville problem, cf.\ \cite{Petrov2004, Citro2008} where this was used to study such models in equilibrium.
As will be seen, this is well suited for our goal as it allows one to readily study non-equilibrium dynamics; this can be contrasted with the complementary approach based on functional integrals as developed in \cite{BrunDubail2018igff}.
Different from the homogeneous case, the eigenmode expansion is not in the usual momentum modes (since translation invariance is explicitly broken, usual momentum is not conserved) but instead in suitably chosen modes encoding the inhomogeneities.

To be more specific, we will show how to diagonalize the Hamiltonian in \eqref{Eq:H_iTLL} when \eqref{vK_condition} is satisfied.
(Recall that this condition implies that the spatial dependence of both the propagation velocity and the Luttinger parameter is given by the same inhomogeneous quantity.)
Without loss of generality we set $K = 1$ in what follows, which can be viewed either as a redefinition $\v \to \v K$ or a rescaling of the fields $\hat{\phi}(x) \to \hat{\phi}(x)/\sqrt{K}$ and $\pif(x) \to \sqrt{K} \pif(x)$, meaning that the effective Hamiltonian can be written as
\begin{equation}
\label{Eq:H_iTLL_vK}
H_{\mathrm{iTLL}}
= \frac{\v}{2} \int_{-R}^{R} \mathrm{d}x\,
    \! \wick{
	    \biggl(
		    \pif(x)^2
		    + F(x) [\partial_{x} \hat{\phi}(x)]^2
    	\biggr)
    } \!,
\end{equation}
where $F(x) = K(x)^2 \geq 0$ and
\begin{equation}
\label{Eq:CCR}
[\hat{\phi}(x), \pif(x')]
= \ii \delta(x - x'),
\qquad
[\hat{\phi}(x), \hat{\phi}(x')]
= 0
= [\pif(x), \pif(x')].
\end{equation}
Taking this as our starting point, it is useful to rewrite the second term in $H_{\mathrm{iTLL}}$ by integrating it by parts.
This yields
$H_{\mathrm{iTLL}}
= H_{F}
  + (\v/2)
    \bigl[
      \! \wick{ \hat{\phi}(x) F(x) \partial_{x} \hat{\phi}(x) } \!
    \bigr]_{x = - R}^{x = R}$
with
\begin{equation}
\label{Eq:H_iTLL_vK_ibp}
H_{F}
= \frac{\v}{2} \int_{-R}^{R} \mathrm{d}x\,
    \! \wick{
	    \biggl(
	    	\pif(x)^2
	    	- \hat{\phi}(x) \partial_{x} [ F(x) \partial_{x} \hat{\phi}(x) ]
	    \biggr)
    } \! .
\end{equation}
The second term above for $H_{\mathrm{iTLL}}$ is a boundary contribution that is assumed fixed by the boundary conditions and will thus be suppressed in what follows.
As such, we will study $H_{F}$, from which it becomes intuitive which boundary conditions to impose and what is the form of the associated Sturm-Liouville operator.

As a remark, a similar discussion to what follows can be performed for effective models that include a tunnel coupling, which in the field-theory description take on the meaning of an inhomogeneous mass term, see Sec.~\ref{Sec:Atom_chip}.
We will not discuss this case here and instead give a brief discussion of these matters in Appendix~\ref{App:iTLL_with_mass_term}.


\subsection{Sturm-Liouville problem}


To diagonalize the Hamiltonian $H_{F}$ we will expand $\hat{\phi}(x)$ and $\pif(x)$ in the eigenfunctions of
\begin{equation}
\mathcal{A} = - \partial_{x} F(x) \partial_{x}
\label{SL_op_A}
\end{equation}
acting on some suitable function space.
We consider the Sturm-Liouville problem
\begin{equation}
\mathcal{A} u
= \lambda u
\end{equation}
for $u$ in the space of twice-differentiable functions satisfying suitable boundary conditions at $x = \pm R$, cf., e.g., \cite{Zettl2010sturm}.
(The boundary conditions are discussed below.)

We recall the important distinction between problems where $F(x) > 0$ for all $x \in [-R, R]$ (including the endpoints) and the case where $F(x)$ can be zero.
Here, if we impose suitable regular boundary conditions, e.g., Dirichlet or Neumann boundary conditions, $F(x) > 0$ implies that our Sturm-Liouville problem is regular, while it is irregular if $F(x) = 0$ somewhere.
Regular problems have a number of general properties that allow one to expand our quantum fields in the eigenfunctions of $\mathcal{A}$.
However, these are known to carry over to the special case in \eqref{K_parabolic} for which $F(\pm R) = 0$.
(This corresponds to, for instance, the vanishing of the mean atom-density distribution at the boundaries of a one-dimensional quasi-condensate.)
We stress that in the latter case, where $F(x)$ is zero at $x = \pm R$ and positive otherwise, the boundary conditions are that any solution must be regular at $x = \pm R$.

For a regular Sturm-Liouville problem, we recall that the eigenvalues are real and can be labeled $\lambda_n$ for $n = 0, 1, \ldots$ so that $0 \leq \lambda_0 < \lambda_1 < \lambda_2 < \ldots$.
One can show that, for each $\lambda_n$, there is a unique smooth eigenfunction $f_n(x)$ that is square integrable and that the set of eigenfunctions form a complete orthonormal basis:
\begin{equation}
\label{ON_basis}
\int_{-R}^{R} \mathrm{d}x\, f_{n\ppr}(x) f_{n'}(x) = \delta_{n, n'},
\qquad
\sum_{n = 0}^{\infty} f_{n}(x) f_{n}(y) = \delta(x - y).
\end{equation}
Moreover, given the above ordering of the eigenvalues, the eigenfunction $f_{n}(x)$ corresponding to $\lambda_{n}$ has $n$ zeroes in the interval $[-R, R]$.  
Thus, even if $F(x)$ varies in space, the eigenfunctions $f_n(x)$ for $n > 0$ are still `oscillatory', qualitatively resembling the cosine eigenfunctions for the case when $F(x)$ is constant (assuming Neumann boundary conditions), while $f_{0}(x) = 1/\sqrt{2R}$ is constant, implying that $\lambda_{0} = 0$.


\subsection{Eigenmode expansion}
\label{Sec:SL_theory:Eigenmode_expansion}


For each mode $n > 0$, consider the Hilbert space of a single harmonic oscillator with creation and annihilation operators denoted by $\hat{a}_{n}^\dagger$ and $\hat{a}_{n}\pdag$, respectively, satisfying $[\hat{a}_{n\ppr}\pdag, \hat{a}_{n'}^{\dagger}] = \delta_{n, n'}$ and $[\hat{a}_{n\ppr}\pdag, \hat{a}_{n'}\pdag] = 0 = [\hat{a}_{n\ppr}^{\dagger}, \hat{a}_{n'}^{\dagger}]$.
The full Hilbert space is constructed as a tensor product of each of these Hilbert spaces together with that for the zero mode, $n = 0$, which requires a separate construction, see, e.g., \cite{Haldane1981, DelftSchoeller1998, LangmannMoosavi2015}:
Let $\QN$ be a self-adjoint operator and $\KF$ a unitary operator that satisfy $[\QN, \KF] = \KF$ and commute with $\hat{a}_{n}^{(\dagger)}$ for all $n > 0$.
The zero-mode part of the full Hilbert space can then be constructed from orthogonal states obtained by $\KF^q$ acting on the vacuum $|\Omega\rangle$ for any integer $q$, where $|\Omega\rangle$ is defined by $\QN |\Omega\rangle = 0 = \hat{a}_n |\Omega\rangle$ for all $n > 0$.
In particular, $\QN \KF^{q} |\Omega\rangle = q \KF^{q} |\Omega\rangle$ and $\langle\Omega| \KF^{q} |\Omega\rangle = \delta_{q, 0}$ for all $q$.
Here, $\QN$ can be interpreted as a charge operator and $\KF$ and $\KF^{\dagger} = \KF^{-1}$ as operators that raise or lower the charge by one,
respectively.

We stress that the separate treatment of zero modes is not particular to inhomogeneous TLLs but is well known and necessary also for homogeneous TLLs.
Indeed, the present discussion covers homogeneous systems as a special case, since the eigenfunctions are general, but the above zero-mode treatment would be unaffected by that simplification.

In terms of $\hat{a}_{n}^\dagger$ and $\hat{a}_{n}\pdag$ we can define
\begin{equation}
\label{phi_n_pi_n}
\hat{\phi}_{n}
= \frac{1}{\sqrt{2}{\lambda_{n}}^{1/4}}
  \left( \hat{a}_{n}\pdag + \hat{a}_{n}^{\dagger} \right),
\qquad
\hat{\pi}_{n}
= - \ii \frac{\lambda_{n}^{1/4}}{\sqrt{2}}
    \left( \hat{a}_{n} - \hat{a}_{n}^{\dagger} \right)
\end{equation}
for $n > 0$, implying $[\hat{\phi}_{n}, \hat{\pi}_{n'}] = \ii \delta_{n, n'}$ and $[\hat{\phi}_{n}, \hat{\phi}_{n'}] = 0 = [\hat{\pi}_{n}, \hat{\pi}_{n'}]$ for $n, n' > 0$.
Moreover, we can define $\hat{\pi}_{0} = \sqrt{\pi/2R} \QN$ and formally identify $\KF$ with $\ee^{ \ii \sqrt{2R/\pi} \hat{\phi}_0 }$; the reason for the latter is that $\hat{\phi}_{0}$ alone is not well defined, see, e.g., \cite{DelftSchoeller1998, LangmannMoosavi2015}.
Their commutation relations follow straightforwardly from the ones above:
It is clear that $\QN$ and $\KF$ commute with $\hat{\phi}_{n}$ and $\hat{\pi}_{n}$ for all $n > 0$, and the definitions are even consistent with $[\hat{\phi}_{n}, \hat{\pi}_{n'}] = \ii \delta_{n, n'}$ formally extended to all $n, n' \geq 0$.

With this we can (formally) write our fields
\begin{equation}
\hat{\phi}(x)
= \frac{\hat{\phi}_{0}}{\sqrt{2R}} + \Delta\hat{\phi}(x),
\qquad
\Delta \hat{\phi}(x)
= \sum_{n = 1}^{\infty} f_{n}(x)\, \hat{\phi}_{n},
\qquad
\pif(x)
= \sum_{n = 0}^{\infty} f_{n}(x)\, \hat{\pi}_{n},
\label{Eq:canquant}
\end{equation}
which then naturally satisfy \eqref{Eq:CCR}.
Here we defined $\Delta\hat{\phi}(x)$ as the field without the zero mode since $\hat{\phi}_{0}$ is not well defined.
One way to motivate studying $\Delta\hat{\phi}(x)$ is to note that $\hat{\phi}(x)$ is only defined up to an overall constant phase; in fact, the more physical quantity is $\partial_x \hat{\phi}(x)$ which is what appears in the Hamiltonian in \eqref{Eq:H_iTLL}, or equivalently, $\hat{\phi}_{0}$ does not appear in this Hamiltonian if expanded in the eigenmodes.
Another motivation is on experimental grounds since relative phases are physically more meaningful and in data analysis of interferometric measurements it is common to reference phase profiles relative to a fixed point in the system.

It is also sometimes useful to (formally) invert the formulas in \eqref{Eq:canquant}:
\begin{equation}
\label{Eq:canquant_inverted}
\hat{\phi}_{n}
= \int_{-R}^{R} \mathrm{d}x\, f_{n}(x) \hat{\phi}(x),
\qquad
\hat{\pi}_{n}
= \int_{-R}^{R} \mathrm{d}x\, f_{n}(x) \pif(x),
\end{equation}
where $\hat{\phi}(x)$ can be replaced by $\Delta\hat{\phi}(x)$ for $n > 0$ since then $\int_{-R}^{R} \mathrm{d}x\, f_{n}(x) = 0$, which follows from \eqref{ON_basis} using that $f_{0}(x)$ is constant.
Lastly, we make precise what the normal ordering $\wick{\cdots}$ entails for multilinears in the fields, namely that all creation operators $\hat{a}_{n}^{\dagger}$ are to the left of all annihilation operators $\hat{a}_{n}\pdag$ when the fields are expanded in eigenmodes.

The above allows us to readily diagonalize the Hamiltonian in \eqref{Eq:H_iTLL_vK_ibp}.
Using \eqref{Eq:canquant} and $\mathcal{A} f_n(x) = \lambda_n f_n(x)$, we obtain
\begin{equation}
\label{H2}
H_{F}
= \frac{\v}{2}
  \left(
    \hat{\pi}_{0}^2
    + \sum_{n = 1}^{\infty}
      \! \wick{
        \left[ \hat{\pi}_{n}^2 + \lambda_{n} \hat{\phi}_{n}^2 \right]
      } \! 
  \right).
\end{equation}
The eigenmodes with $n>0$ enter as harmonic-oscillator modes with energies 
\begin{equation}
\label{E_n}
E_{n} = \v\sqrt{\lambda_{n}}.
\end{equation}
Substituting this into \eqref{H2} we can express the Hamiltonian in diagonal form as
\begin{equation}
\label{H_F_diagonalized}
H_{F}
= \frac{\v}{2} \hat{\pi}_0^2
  + \sum_{n = 1}^{\infty} E_{n}
    \hat{a}_{n}^{\dagger} \hat{a}_{n}\pdag.
\end{equation}
This shows that $|\Omega\rangle$ is the ground state of $H_{F}$, and thus also of $H_{\mathrm{iTLL}}$.

In what follows we will be interested in the irregular Sturm-Liouville problem given by \eqref{K_parabolic}, i.e., $F(x) = 1 - (x/R)^{2}$.
The solutions to the corresponding eigenvalue problem are
\begin{equation}
\label{iTLL_parabolic_lambda_n_f_n}
\lambda_{n}
= \frac{n(n+1)}{R^2},
\qquad
f_{n}(x)
= \sqrt{\frac{2n+1}{2R}} P_{n}(x/R),
\quad
n = 0, 1, \dots,
\end{equation}
where $P_{n}(\cdot)$ are Legendre polynomials.
(The eigenfunctions satisfy the imposed boundary conditions since they are regular at $\pm R$.)
As mentioned, one can show that the properties that we have listed above for regular Sturm-Liouville problems also hold for this irregular special case.


\section{Dynamics for inhomogeneous TLLs}
\label{Sec:Dynamics_for_iTLLs}


In what follows we use the tools in Sec.~\ref{Sec:SL_theory} to compute results describing the dynamics of inhomogeneous TLLs.


\subsection{Correlation functions}
\label{Sec:Dynamics_for_iTLLs:CF}

\begin{figure}[!htbp]

\centering

\subfigure[Inhomogeneous TLL]{%
\includegraphics[height=0.4\linewidth]{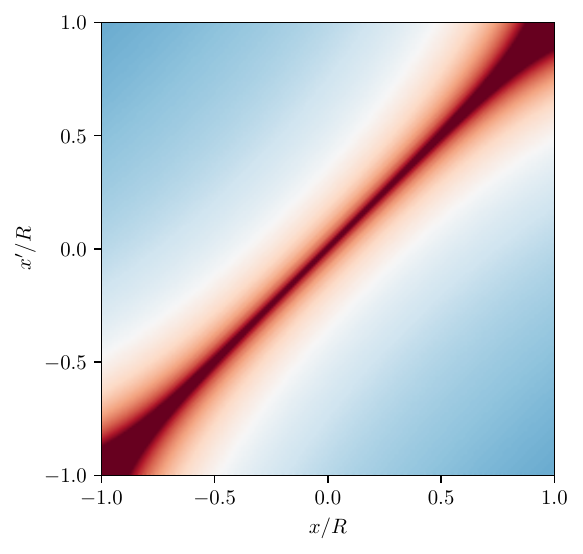}
\label{Fig:F_GS_xt_xptp:iTLL}%
}%
\hspace{10mm}
\subfigure[Inhomogeneous-velocity wave equation]{%
\includegraphics[height=0.4\linewidth]{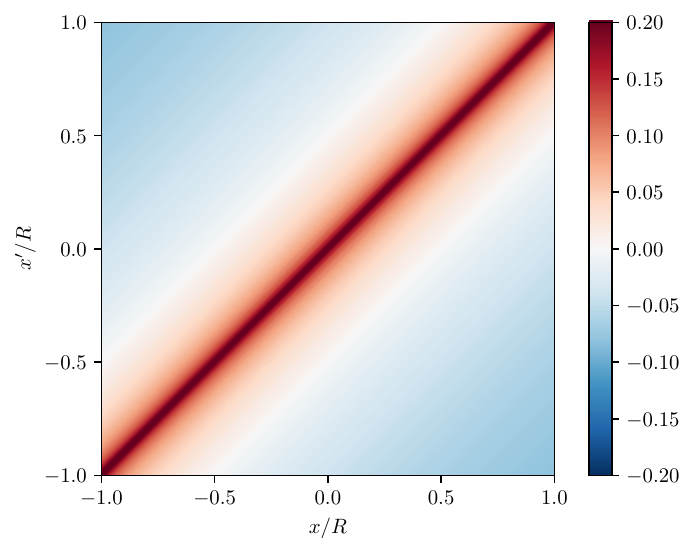}
\label{Fig:F_GS_xt_xptp:iVWE}%
}%

\caption{%
Plots of the equal-time ground-state correlation function $F_{\mathrm{GS}}(x, x'; 0)$ in \eqref{F_GS_xt_xptp} for \subref{Fig:F_GS_xt_xptp:iTLL} inhomogeneous TLL with $E_{n}$ and $f_{n}(x)$ in \eqref{E_n} and \eqref{iTLL_parabolic_lambda_n_f_n} and \subref{Fig:F_GS_xt_xptp:iVWE} the corresponding inhomogeneous-velocity wave equation (see Sec.~\ref{Sec:Comparison_iVWE}).
(To regularize the plotted results, a smooth ultraviolet cutoff is used.)%
}%

\label{Fig:F_GS_xt_xptp}

\end{figure}

As examples of descriptive quantities of the dynamics we begin by computing non-equal-time correlation functions.
For instance, using \eqref{phi_n_pi_n} and \eqref{Eq:canquant}, one can show that the two-point correlation function for $\Delta\hat{\phi}(x, t) = \ee^{\ii H_{\mathrm{iTLL}} t} \Delta\hat{\phi}(x) \ee^{-\ii H_{\mathrm{iTLL}} t}$ in the ground state $|\Omega\rangle$ of $H_{\mathrm{iTLL}}$ is
\begin{equation}
\label{F_GS_xt_xptp}
F_{\mathrm{GS}}(x, x'; t - t')
= \langle \Delta\hat{\phi}(x, t) \Delta\hat{\phi}(x', t') \rangle_{\mathrm{GS}}
= \sum_{n = 1}^{\infty} \frac{\v}{2E_{n}}
  f_{n}(x) f_{n}(x') \ee^{-\ii E_{n} (t - t') - n0^+},
\end{equation}
where $\langle \cdots \rangle_{\mathrm{GS}} = \langle\Omega| \cdots |\Omega\rangle$ and we inserted an exponential cutoff as a prescription for making the sum well defined.
Recall that $\Delta\hat{\phi}(x)$ does not contain the zero mode and that this can be motivated both on mathematical and experimental grounds, see \eqref{Eq:canquant} and the accompanying discussion.
The result in \eqref{F_GS_xt_xptp} is plotted in Fig.~\ref{Fig:F_GS_xt_xptp} for the special case in \eqref{iTLL_parabolic_lambda_n_f_n} corresponding to \eqref{K_parabolic} at equal time $t = t'$.

Other such examples are non-equal-time correlation functions for vertex operators.
These are important since, in the harmonic-fluid approach, the bosonic or fermionic fields of any original model are expressed using such operators \cite{Haldane1981, Cazalilla2004, Giamarchi2003}.
In our case, these vertex operators are of the form (we omit higher harmonics for simplicity)
\begin{equation}
\label{V_wab}
V_{w, a, b}(x)
= \wick{
    \KF^{w}
    \ee^{
      \ii a \sqrt{\pi} \hat{\theta}(x)
      + \ii b \sqrt{\pi} \Delta\hat{\phi}(x)
    }
  }
\end{equation}
for integers $w$ and real numbers $a$ and $b$, 
where we introduced $\hat{\theta}(x)$ so that $\partial_x \hat{\theta}(x) = \pif(x)$, which implies $[\hat{\phi}(x), \hat{\theta}(x')] = (\ii/2) \sgn(x-x')$.
The absence of a factor $\pi$ on the right-hand side of the last relation compared to \cite{Haldane1981, Cazalilla2004} explains the inclusions of the factors $\sqrt{\pi}$ in \eqref{V_wab}: Our fields $\hat{\phi}(x)$ and $\hat{\theta}(x)$ differ by exactly this factor from the corresponding fields in \cite{Haldane1981, Cazalilla2004}.
We stress that, in concrete applications, $w$, $a$, and $b$ would depend non-trivially on parameters of the original model, but we keep them unspecified in what follows.

Three remarks:
(i) Expanded in eigenmodes, we (formally) have
\begin{equation}
\hat{\theta}(x)
= \sqrt{\pi} x \frac{\QN}{2R} + \sum_{n = 1}^{\infty} F_{n}(x) \hat{\pi}_{n},
\qquad
F_{n}(x) = \int_{0}^{x} \mathrm{d}x' f_{n}(x') \quad (n = 1,2, \ldots),
\end{equation}
where we used $f_{0}(x) = 1/\sqrt{2R}$ and $\hat{\pi}_0 = \sqrt{\pi/2R} \QN$ for the zero-mode part.
(ii) We recall that $\wick{\cdots}$ denotes (boson) normal ordering and refer to \cite{DelftSchoeller1998, LangmannMoosavi2015} for what this entails for vertex operators, namely that
$\wick{ \KF^w \ee^{\ii \pi ax \QN/2R} }
= \ee^{\ii \pi ax \QN/4R} \KF^w \ee^{\ii \pi ax \QN/4R}$
for the zero modes
and
$\wick{ \ee^{\ii \sum_{n > 0} (c_{n}\pdag \hat{a}_{n}^{\dagger} + d_{n}\pdag \hat{a}_{n}\pdag)} }
= \ee^{\ii \sum_{n > 0} c_{n}\pdag \hat{a}_{n}^{\dagger}} \ee^{\ii \sum_{n > 0} d_{n}\pdag \hat{a}_{n}\pdag}$
($c_{n}, d_{n} \in \mathbb{C}$) for the non-zero modes.
(iii) It is important to treat $\KF$ and $\Delta\hat{\phi}(x,t)$ separately as only integer powers of the former are well defined, see \eqref{Eq:canquant} and the introduction of $\KF$ in Sec.~\ref{Sec:SL_theory:Eigenmode_expansion}.

We are interested in general non-equal-time ground-state correlation functions of the form
\begin{equation}
C_{\mathrm{GS}}(x, x'; t - t')
= \bigl\langle
    V_{w, a, b}(x, t)^{\dagger}
    V_{w, a, b}(x', t')
  \bigr\rangle_{\mathrm{GS}},
\end{equation}
where $V_{w, a, b}(x, t) = \ee^{\ii H_{\mathrm{iTLL}} t} V_{w, a, b}(x) \ee^{-\ii H_{\mathrm{iTLL}} t}$.
The result is
\begin{align}
\label{C_xt_xptp_def_result_w_a_b}
C_{\mathrm{GS}}(x, x'; t - t')
& =
  \exp
  \left(
    - \frac{\ii \pi w}{2} \frac{a (x-x') + w v (t-t')}{2R}
  \right) \nonumber \\
& \quad \times
  \exp
  \left(
    \sum_{n = 1}^{\infty}
    \frac{\pi v}{2E_{n}}
    \left[
      bf_{n}(x) - \ii a \frac{E_{n}}{v} F_{n}(x)
    \right]
    \left[
      bf_{n}(x') + \ii a \frac{E_{n}}{v} F_{n}(x')
    \right]
    \ee^{-\ii E_{n}(t-t') - n 0^+}
  \right),
\end{align}
where we again inserted an exponential cutoff as a prescription for making the sum well defined.
The expression above can be obtained using a special case of the Baker-Campbell-Hausdorff formula and other standard manipulations for vertex operators, see, e.g., \cite{DelftSchoeller1998, LangmannMoosavi2015} for such details.
We note that \eqref{C_xt_xptp_def_result_w_a_b} can be shown to reduce to known expressions for special cases, such as a homogeneous TLL in a box \cite{Cazalilla2004}.

\begin{figure}[!htbp]

\centering

\subfigure[Inhomogeneous TLL]{%
\includegraphics[width=0.99\linewidth]{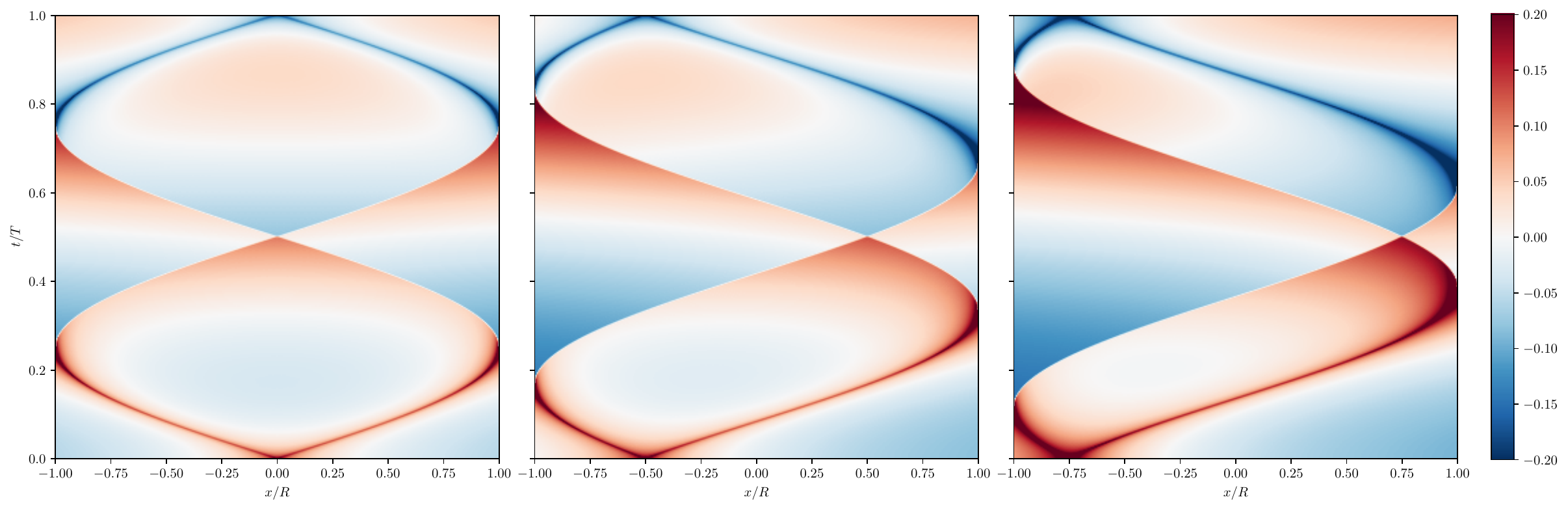}
\label{Fig:Re_F_GS_xt_xptp:iTLL}
}%

\subfigure[Inhomogeneous-velocity wave equation]{%
\includegraphics[width=0.99\linewidth]{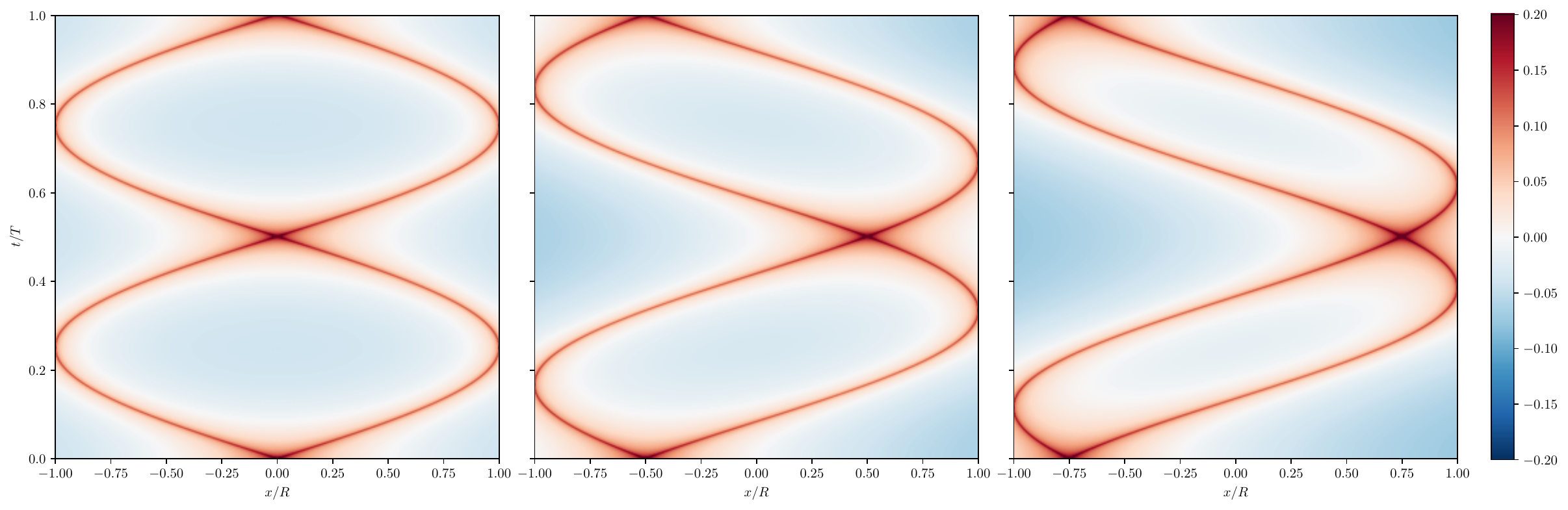}
\label{Fig:Re_F_GS_xt_xptp:iVWE}
}%

\caption{%
Density plots of $\Re F_{\mathrm{GS}}(x_{0}, x; t)$ in \eqref{F_GS_xt_xptp} for the non-equal-time correlation function in \eqref{C_xt_xptp_def_result_w0_a0} as a function of $x$ and $t$ for $x_0 = 0$, $-R/2$, and $-3R/4$ (left to right) for
\subref{Fig:Re_F_GS_xt_xptp:iTLL} inhomogeneous TLL with $E_{n}$ and $f_{n}(x)$ in \eqref{E_n} and \eqref{iTLL_parabolic_lambda_n_f_n} and
\subref{Fig:Re_F_GS_xt_xptp:iVWE} the corresponding inhomogeneous-velocity wave equation (see Sec.~\ref{Sec:Comparison_iVWE}).
As in Fig.~\ref{Fig:Propagation_of_Gaussian_wp_introduction}, the plots in \subref{Fig:Re_F_GS_xt_xptp:iTLL} exhibit a sign inversion after one (quasi-)period $T = 2\pi R/v$ while those in \subref{Fig:Re_F_GS_xt_xptp:iVWE} do not.
(To regularize the plotted results, a smooth ultraviolet cutoff is used.)%
}%

\label{Fig:Re_F_GS_xt_xptp}

\end{figure}

In the simpler cases when $w = a = 0$ or $w = b = 0$, the correlation functions for the vertex operators can be expressed in terms of the correlations functions for the fields.
For instance,
\begin{equation}
\left| \bigl\langle
    \wick{ \ee^{-\ii b \sqrt{\pi} \Delta\hat{\phi}(x, t)} }
    \!\!
    \wick{ \ee^{\ii b \sqrt{\pi} \Delta\hat{\phi}(x', t')} }
  \bigr\rangle_{\mathrm{GS}} \right|
= \exp
  \Bigl(
    \pi b^{2} \Re F_{\mathrm{GS}}(x, x'; t - t')
  \Bigr)
\label{C_xt_xptp_def_result_w0_a0}
\end{equation}
using $F_{\mathrm{GS}}(x, x'; t - t')$ in \eqref{F_GS_xt_xptp}.
The real part of the latter is plotted in Fig.~\ref{Fig:Re_F_GS_xt_xptp} for the special case in \eqref{iTLL_parabolic_lambda_n_f_n} corresponding to \eqref{K_parabolic}.

We note that, while the plots with or without a position-dependent $K(x)$ look similar in Fig.~\ref{Fig:F_GS_xt_xptp} at equal time $t = t'$, the dynamics shown in Fig.~\ref{Fig:Re_F_GS_xt_xptp} reveal a striking difference, as discussed in detail below.


\subsection{Green's functions}
\label{Sec:Dynamics_for_iTLLs:GF}


To better understand the dynamics of inhomogeneous TLLs, such as the propagation of initial wave packets evolving under $H_{\mathrm{iTLL}}$, we also use our tools to derive Green's functions for the Heisenberg equations of motion.

For instance, from the Heisenberg equation using $H_{\mathrm{iTLL}}$ in \eqref{Eq:H_iTLL_vK} [or equivalently, $H_{F}$ in \eqref{Eq:H_iTLL_vK_ibp}] and the commutation relations in \eqref{Eq:CCR} we (formally) obtain
\begin{equation}
\label{dt_phi_dt_pi}
\partial_{t} \hat{\phi}
= \v \pif,
\qquad
\partial_{t} \pif
= \v \partial_{x} \left[ F(x) \partial_{x} \hat{\phi} \right].
\end{equation}
This implies that the equation of motion for $\hat{\phi}$ is
\begin{equation}
\label{EoM_phi}
\partial_{t}^{2} \hat{\phi}
= \v^{2} \partial_{x} \left[ F(x) \partial_{x} \hat{\phi} \right],
\end{equation}
which can be seen as a certain inhomogeneous generalization of the wave equation.
Its general solution (for arbitrary initial conditions) can be found by expanding $\hat{\phi}(x, t)$ in the eigenfunctions using \eqref{phi_n_pi_n} and \eqref{Eq:canquant}:
\begin{equation}
\hat{\phi}_{n}(t)
= \sqrt{\frac{\v}{2E_{n}}}
  \left(
    \hat{a}_{n} \ee^{-\ii E_{n}t}
    + \hat{a}_{n}^{\dagger} \ee^{\ii E_{n}t}
  \right)
= \hat{\phi}_{n} \partial_{t} g_{n}(t) + \v \hat{\pi}_{n} g_{n}(t)
\end{equation}
with
\begin{equation}
\label{g_n_t}
g_{n}(t) = \frac{\sin(E_{n}t)}{E_{n}}.
\end{equation}
As argued before, we will consider $\Delta\hat{\phi}(x, t)$ without the zero mode, see \eqref{Eq:canquant}.
From the above together with \eqref{Eq:canquant_inverted} we find that the general solution in position space is
\begin{equation}
\Delta\hat{\phi}(x, t)
= \int_{-R}^{R} \mathrm{d}x'\,
  \left(
    G_{\phi\phi}(x, x'; t)\, \Delta\hat{\phi}(x', 0)
    + G_{\phi\pi}(x, x'; t)\, \pif(x', 0)
  \right),
\end{equation}
where
\begin{equation}
\label{G_phi_phi_G_phi_pi}
G_{\phi\phi}(x, x'; t)
= \sum_{n = 1}^{\infty} f_{n}(x) f_{n}(x') \partial_{t} g_{n}(t),
\qquad
G_{\phi\pi}(x, x'; t)
= \v \sum_{n = 1}^{\infty} f_{n}(x) f_{n}(x') g_{n}(t)
\end{equation}
are the relevant Green's functions.
Note that $G_{\phi\pi}(x, x'; 0) = 0$ while $G_{\phi\phi}(x, x'; 0)$ equals $\delta(x-x')$ due to \eqref{ON_basis} up to a constant term that precisely corresponds to the zero-mode term excluded from $\Delta\hat{\phi}(x)$.

Recalling that, to linear approximation within the harmonic-fluid approach, $\pif(x)$ and $v(x) K(x) \partial_{x} \hat{\phi}(x)$ can be interpreted as the particle density and current, respectively, cf.\ the second formula in \eqref{dt_phi_dt_pi}, it is instructive to also compute their Green's functions.
To simplify notation, we write $\partial\hat{\phi}(x) = \partial_{x} \hat{\phi}(x)$.
In Appendix~\ref{Appendix:GF} we show that
\begin{equation}
\label{dphi_GG}
\partial\hat{\phi}(x, t)
= \int_{-R}^{R} \mathrm{d}x'\,
  \left(
    G_{\partial\phi\partial\phi}(x, x'; t)\, \partial\hat{\phi}(x', 0)
    + G_{\partial\phi\pi}(x, x'; t)\, \pif(x', 0)
  \right)
\end{equation}
with
\begin{equation}
\label{G_dphi_dphi_G_dphi_pi}
G_{\partial\phi\partial\phi}(x, x'; t)
= - \int_{-R}^{x'} \mathrm{d}x''\, \partial_{x} G_{\phi\phi}(x, x''; t),
\qquad
G_{\partial\phi\pi}(x, x'; t)
= \partial_{x} G_{\phi\pi}(x, x'; t).
\end{equation}
The corresponding Green's functions for $\pif(x, t) = \ee^{\ii H_{\mathrm{iTLL}} t} \pif(x) \ee^{-\ii H_{\mathrm{iTLL}} t}$ are given in Appendix~\ref{Appendix:GF}.

\begin{figure}[!htbp]

\centering

\subfigure[$G_{\partial\phi\partial\phi}(x, x'; t)$]{
\includegraphics[width=0.95\linewidth]{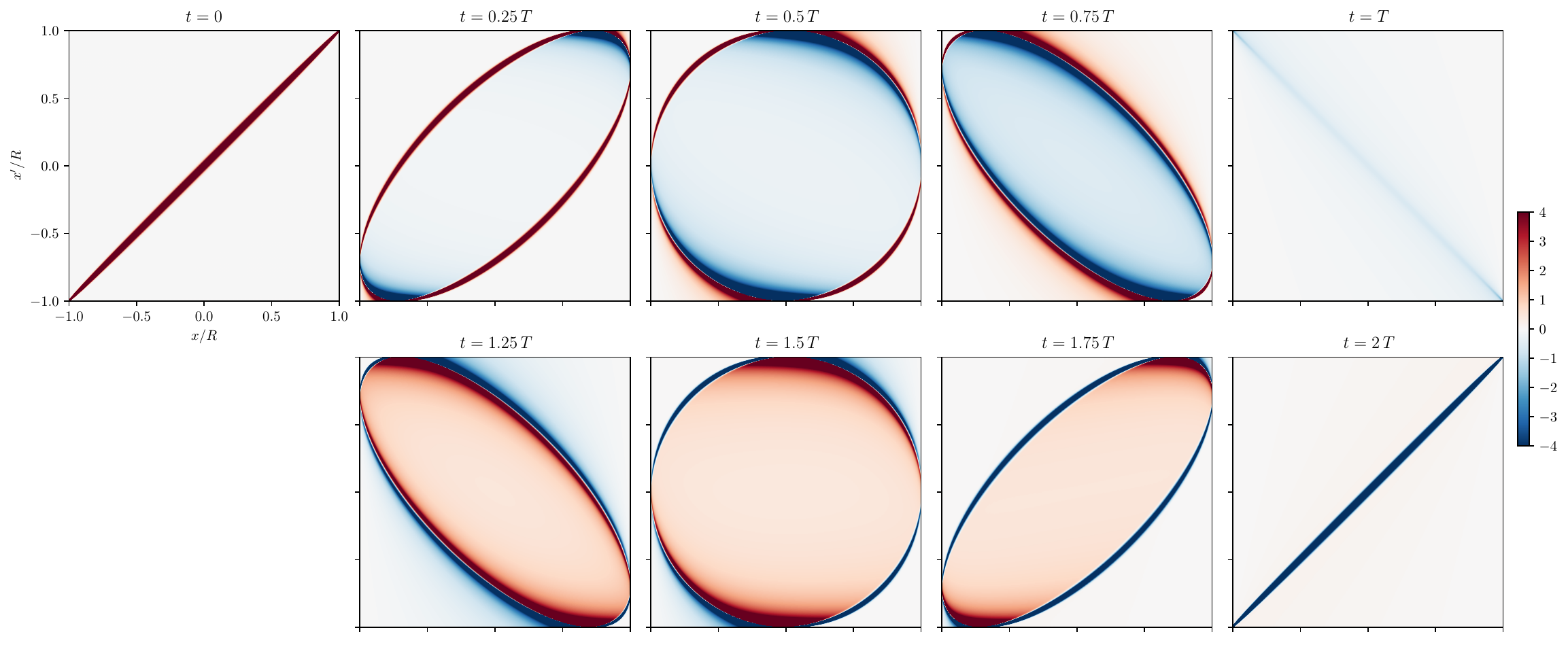}
\label{Fig:Green_functions_dphi:G_dphi_dphi}
}

\subfigure[$G_{\partial\phi\pi}(x, x'; t)$]{
\includegraphics[width=0.95\linewidth]{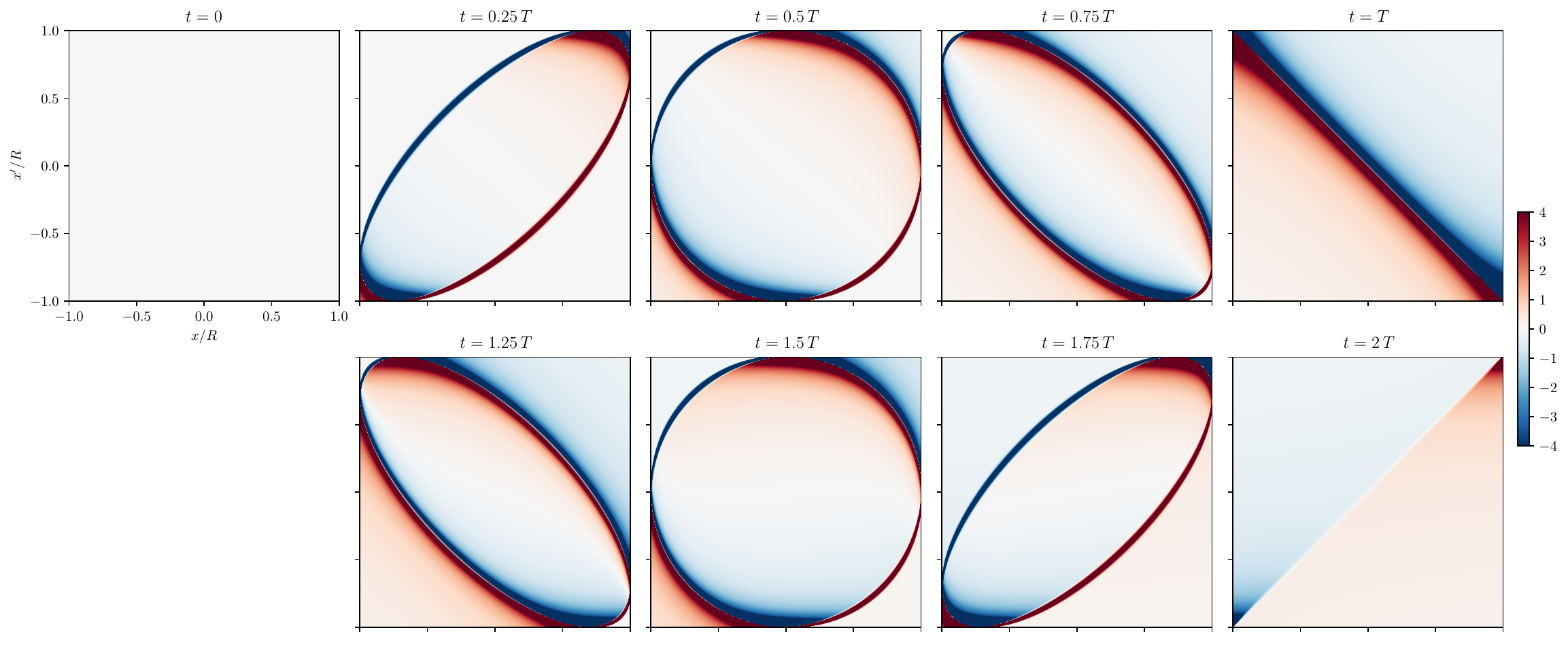}
\label{Fig:Green_functions_dphi:G_dphi_pi}
}

\caption{%
Green's functions
\subref{Fig:Green_functions_dphi:G_dphi_dphi} $G_{\partial\phi\partial\phi}(x, x'; t)$
and
\subref{Fig:Green_functions_dphi:G_dphi_pi}
$G_{\partial\phi\pi}(x, x'; t)$
in \eqref{G_dphi_dphi_G_dphi_pi} for inhomogeneous TLL with $E_{n}$ and $f_{n}(x)$ in \eqref{E_n} and \eqref{iTLL_parabolic_lambda_n_f_n}
for times $t/T = 0$, $1/4$, $1/2$, $\dots$, $2$, where $T = 2 \pi R/\v$ is the period of approximate recurrences.
(To regularize the plotted results, a smooth ultraviolet cutoff is used.)%
}%

\label{Fig:Green_functions_dphi}

\end{figure}

\begin{figure}[!htbp]

\centering

\subfigure[$G_{\partial\phi\partial\phi}(0, x; t)$]{%
\includegraphics[height=0.4\linewidth]{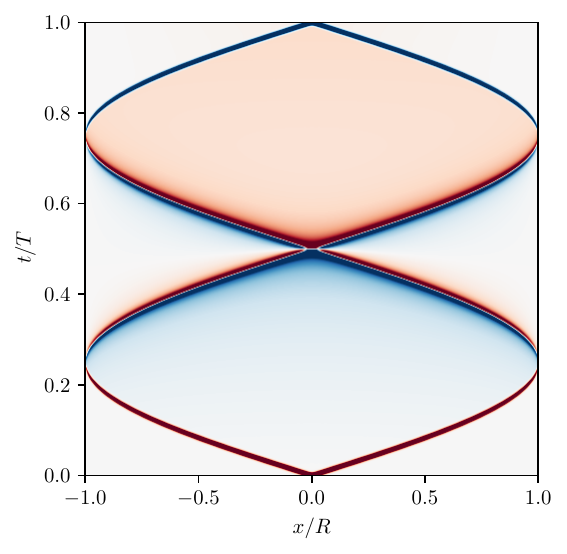}
\label{Fig:Green_functions_dphi_timeG_dphi_dphi}
}%
\hspace{10mm}
\subfigure[$G_{\partial\phi\pi}(0, x; t)$]{%
\includegraphics[height=0.4\linewidth]{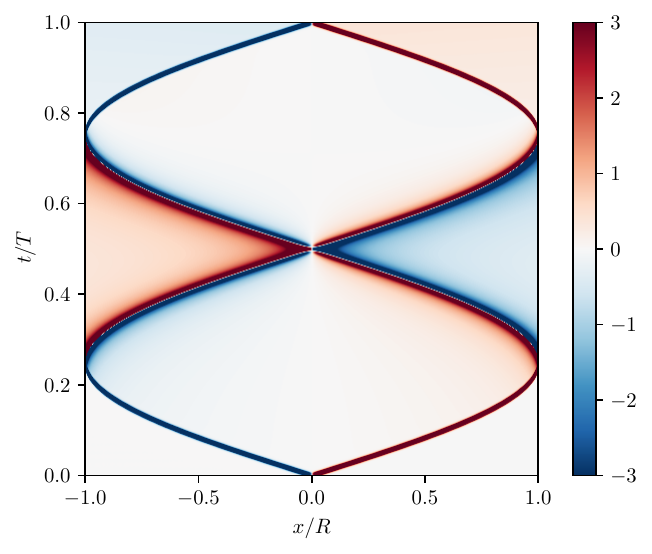}
\label{Fig:Green_functions_dphi_timeG_dphi_pi}
}%

\caption{%
Green's functions \subref{Fig:Green_functions_dphi_timeG_dphi_dphi} $G_{\partial\phi\partial\phi}(0, x; t)$ and  \subref{Fig:Green_functions_dphi_timeG_dphi_pi} $G_{\partial\phi\pi}(0, x; t)$ as functions of $x$ and $t$ for the same case as in Fig.~\ref{Fig:Green_functions_dphi}.%
}%

\label{Fig:Green_functions_dphi_time}

\end{figure}

The Green's functions $G_{\partial\phi\partial\phi}(x, x'; t)$ and $G_{\partial\phi\pi}(x, x'; t)$ are plotted in Figs.~\ref{Fig:Green_functions_dphi} and~\ref{Fig:Green_functions_dphi_time} for the special case in \eqref{iTLL_parabolic_lambda_n_f_n} corresponding to \eqref{K_parabolic}.
From these plots we observe that the propagation given by \eqref{EoM_phi} has the following characteristic properties:
\begin{enumerate}[label={(\alph*)}, ref={(\alph*)}]

\item
\label{Property:CLC}
Curved light cones:
The wavefronts do not follow straight but curved spacetime trajectories given by the equation
$[\arccos(x/R) - \arccos(x'/R)]^2 = (\v t/R)^2$.
Causality is satisfied, i.e., $G(x, x', t) = 0$ for all points $x$ and $x'$ outside these fronts at a given time $t$ (cf.\ the top left and bottom right  corners of the first four time slices in Fig.~\ref{Fig:Green_functions_dphi}).

\item
\label{Property:HFP}
Ballistically non-localized dynamics:
The Green's functions exhibit long tails behind the wavefronts, whose support extends through the entire region enclosed by them, implying that the Huygens-Fresnel principle is not valid.

\item
\label{Property:AR}
Approximate recurrences:
The dynamics exhibits behavior with strongly pronounced approximate recurrences with period $T = 2\pi R/\v$.
The latter is the time needed for wavefronts to reach the boundaries and return to their origin for the first time.

\end{enumerate}

Properties~\ref{Property:CLC} and~\ref{Property:AR} can be demonstrated for the special case in \eqref{K_parabolic} from the asymptotics of the energy eigenvalues $E_n$ and the eigenfunctions $f_n(x)$ in \eqref{E_n} and \eqref{iTLL_parabolic_lambda_n_f_n} for large $n$.
Indeed, the large $n$ asymptotics of Legendre polynomials \cite[\href{https://dlmf.nist.gov/18.15.iii}{{\S}18.15(iii)}]{NIST:DLMF} says that
\begin{equation}
P_{n}(\cos\theta)
\sim
  \frac{2}{\sqrt{2\pi n\sin\theta}} 
  \cos([n + 1/2] \theta - \pi/4),
\label{Legendre_asympt}
\end{equation}
while \eqref{E_n} and \eqref{iTLL_parabolic_lambda_n_f_n} imply
\begin{equation}
E_{n} \sim \frac{(n + 1/2) \v}{R}
\label{E_asympt}
\end{equation}
for large $n$.
From the above relations we find, e.g., that the contribution to $G_{\phi\phi}(x, x'; t)$ from high-energy modes is
\begin{align}
G_{\phi\phi}(x, x'; t)
& \sim
    \sum_{n = 1}^{\infty}
    \frac{
      \cos \bigl( [n+1/2] \arccos(x/R) - \pi/4 \bigr)
      \cos \bigl( [n+1/2] \arccos(x'/R) - \pi/4 \bigr)
      \cos \bigl( [n+1/2] \v t/R \bigr)
    }{
      \pi [1 - (x/R)^{2}]^{1/4} [1 - (x'/R)^{2}]^{1/4}
    }
    \nonumber \\
& = \frac{
      \delta \bigl( \arccos(x/R) - \arccos(x'/R) - \v t/R \bigr)
      + \delta \bigl( \arccos(x/R) - \arccos(x'/R) + \v t/R \bigr)
    }{
      2 [1 - (x/R)^{2}]^{1/4} [1 - (x'/R)^{2}]^{1/4}
    },
\end{align}
and using \eqref{G_dphi_dphi_G_dphi_pi} one can obtain the corresponding result for $G_{\partial\phi\partial\phi}(x, x'; t)$.
These kinds of results for $G_{\phi\pi}(x, x'; t)$ and $G_{\partial\phi\pi}(x, x'; t)$ follow similarly.
The above shows that the dynamics exhibits two oppositely moving fronts that propagate according to
\begin{equation}
\arccos(x/R) - \arccos(x'/R) = \pm \v t/R.
\label{lightcone1}
\end{equation}
Moreover, approximate recurrences occur at integer multiples of $T = 2\pi R/\v$, which follows from the asymptotic linearity of the dispersion relation in \eqref{E_asympt}.
Nevertheless, it is interesting to note that, asymptotically, the energy levels are not integer multiples of the frequency $\omega = \v/R$ but half-integer multiples, which results in an approximate $\pi$ phase shift in the time-evolution operator over a quasi-period $T$.
This is manifested in Figs.~\ref{Fig:Propagation_of_Gaussian_wp_introduction},~\ref{Fig:Re_F_GS_xt_xptp}, and~\ref{Fig:Green_functions_dphi_time} by the presence of a sign inversion after one quasi-period $T$ for inhomogeneous TLL.

While Properties~\ref{Property:CLC} and~\ref{Property:AR} can be seen from the asymptotic properties of the eigenfunctions and eigenvalues at high energy (large $n$), Property~\ref{Property:HFP} is linked to the non-linearity of the dispersion relation at low energy (small $n$).
As we observe in Figs.~\ref{Fig:Green_functions_dphi} and~\ref{Fig:Green_functions_dphi_time}, the Green's functions $G_{\partial\phi\partial\phi}(x, x'; t)$ and $G_{\partial\phi\pi}(x, x'; t)$ spread through the entire space enclosed by the wavefronts and also behind the inwards-moving wavefronts that are reflected off the boundaries of the system, see also Figs.~\ref{Fig:Propagation_of_Gaussian_wp_introduction} and~\ref{Fig:Re_F_GS_xt_xptp}.
The magnitude of their values inside the light cone starts from zero and increases with time.
This behavior is indeed what we would expect from a system with non-linear dispersion, unlike standard TLL or its inhomogeneous-velocity version (with constant Luttinger parameter).


\section{Comparison with wave equation with an inhomogeneous velocity}
\label{Sec:Comparison_iVWE}


The equation in \eqref{EoM_phi} is different from the equation of motion corresponding to a TLL where only the velocity $v(x)$ is inhomogeneous while the Luttinger parameter is constant. 
The latter is described by the inhomogeneous-velocity wave equation
\begin{equation}
\partial_{t}^{2} \hat{\phi}
= v(x)\partial_{x} \left[ v(x)\partial_{x} \hat{\phi} \right]
= v(x)v'(x) \partial_{x} \hat{\phi}
  + v(x)^2\partial^2_{x} \hat{\phi}.
\label{iVWE}
\end{equation}
On the contrary, \eqref{EoM_phi} can equivalently be written in the comparable form
\begin{equation}
\partial_{t}^{2} \hat{\phi}
= 2 v(x) v'(x) \partial_{x} \hat{\phi}
  + v(x)^2 \partial^2_{x} \hat{\phi},
\label{EoM_phi_ver2}
\end{equation}
since $F(x) = K(x)^2 = [v(x)/\v]^2$ in the case we consider here.
The presence of the factor $2$ in the first term above is crucial: It means that \eqref{EoM_phi_ver2} cannot be rewritten as a standard wave equation by changing coordinates, different from \eqref{iVWE}.
Indeed, by a change of variables to $y = f(x)$, where
\begin{equation}
\label{f_x}
f(x) = \int_{0}^{x} \mathrm{d}x'\, \frac{\vnull}{v(x')},
\qquad
\frac{1}{\vnull}
= \frac{1}{2R} \int_{-R}^{R} \mathrm{d}x'\, \frac{1}{v(x')},
\end{equation}
we obtain $v(x) \partial_{x} = \vnull \partial_{y}$, which together with defining $\tilde{\phi}(y, t) = \hat{\phi}(x, t)$ implies that \eqref{iVWE} becomes
\begin{equation}
\partial_{t}^{2} \tilde{\phi}
= \vnull^{2} \partial_{y}^{2} \tilde{\phi}.
\end{equation}

Assuming Neumann boundary conditions for $\hat{\phi}$ at $x = \pm R$, these translate into $(\partial_{y} \tilde{\phi})|_{y = f(\pm R)} = 0$ for the reformulated problem.
Thus, expressed using $y$, the eigenvalues and the eigenfunctions corresponding to \eqref{iVWE} are those of a homogeneous wave equation on the interval of the same length $f(R) - f(-R) = 2R$ centered at $y_0 = [f(R) + f(-R)]/2$ with Neumann boundary conditions and wave velocity $\vnull$:
\begin{equation}
\label{iVWE_tE_n_tf_n}
\tilde{E}_{n}
= \frac{n\pi \vnull}{2R},
\qquad
\tilde{f}_{n}(y)
= \begin{cases}
    \frac{1}{\sqrt{2R}}
    & \quad \text{if } n = 0, \\
    \frac{1}{\sqrt{R}} \cos(n\pi/2 + n\pi[y - y_0]/2R)
    & \quad \text{if } n = 1, 2, \ldots
  \end{cases}
\end{equation}
with $y = f(x)$ given by \eqref{f_x}.
Results involving $\hat{\phi}$ can thus be obtained using the above eigenfunctions and changing coordinates from $y$ to $x$.
Similarly for vertex operators constructed using $\hat{\phi}$, results follow using transformation rules for such vertex operators under a coordinate transformation.

For the dynamics given by \eqref{iVWE}, our special case [cf.\ \eqref{K_parabolic}] corresponds to $v(x) = \v \sqrt{1 - (x/R)^{2}}$, which implies $\vnull = 2\v/\pi$, $y_{0} = 0$, and
\begin{equation}
f(x) = \frac{2R}{\pi} \arcsin(x/R).
\end{equation}
In this special case, even though the eigenfunctions in \eqref{iVWE_tE_n_tf_n} are not exactly the same as the asymptotic form given by \eqref{Legendre_asympt} for those of an inhomogeneous TLL with $K(x)$ and $v(x)$ satisfying \eqref{K_parabolic}, they give rise to exactly the same curved light-cone trajectories:
\begin{equation}
\arcsin(x/R) - \arcsin(x'/R) = \pm \v t/R.
\label{lightcone2}
\end{equation}
These can be seen to be equivalent to those in \eqref{lightcone1}.

Moreover, from \eqref{iVWE_tE_n_tf_n} we see that recurrences
are also present here, as for an inhomogeneous TLL in the case of a parabolic trap.
However, unlike in that case, the recurrences of \eqref{iVWE} with $v(x) = \v \sqrt{1 - (x/R)^{2}}$ are exact instead of approximate (since $\tilde{E}_{n} = n\v/R$ always), with recurrence period $T = 4R/\vnull = 2\pi R/\v$ and there is no sign inversion, cf.\ Figs.~\ref{Fig:Propagation_of_Gaussian_wp_introduction},~\ref{Fig:Re_F_GS_xt_xptp}, and~\ref{Fig:Green_functions_dphi_time} for the inhomogeneous-velocity wave equation.

The above two properties (cf.\ Properties~\ref{Property:CLC} and~\ref{Property:AR} in Sec.~\ref{Sec:Dynamics_for_iTLLs:GF}) shared by the two different models are expected to be more generally valid on the basis of the WKB approximation:
Applied to either the eigenvalue equations or the equations of motion, it shows that the high-energy asymptotics are the same for inhomogeneous TLLs and the inhomogeneous-velocity wave equation, since the dominant behavior is determined by the highest-derivative terms of the equations of motion [cf.\ \eqref{iVWE} and \eqref{EoM_phi_ver2}].
The main difference between \eqref{EoM_phi} and \eqref{iVWE} is, however, that the Huygens-Fresnel principle is valid for the latter but not for the former. 


\section{Possible experimental observations in ultracold atoms}
\label{Sec:Experimental_observations}


In this section we discuss possibilities of experimentally observing the signatures of breaking of the Huygens-Fresnel principle in ultracold atoms by studying first moments of density fluctuations.
Intuitively this is done by looking from the side; more specifically, in the experiment, this involves illuminating the gas with a light field and observing the density absorption by the gas.
This allows one to measure the single shot distribution of the atom density as a function of space and time, cf.\ \cite{Rauer2018}.
Using this technique brings us to the following proposal for studying the Huygens-Fresnel principle in a cold-atom experiment:

Consider a one-dimensional gas trapped on an atom chip with an inhomogeneous mean atom-density distribution.
By repeating the experiment, one can estimate this distribution $\nGP(x)$ using an average over realizations.
Thus, we can verify that the system loaded into the trapping potential and cooled to thermal equilibrium realizes the desired reference distribution, which gives rise to $v(x)$ and $K(x)$ in the inhomogeneous TLL Hamiltonian.
By averaging over realizations, we would additionally find that the first moment of the density-fluctuation field vanishes up to statistical fluctuations.
Using the digital micromirror device (DMD) functionality available on the atom chip~\cite{Tajik2019}, one can additionally, at the initial time, shine a localized repulsive light field onto the atoms.
This should induce a density distribution having a value that agrees with the desired reference density distribution outside the region illuminated by the local light field and a non-zero modification inside that region.
At $t = 0$, we suddenly switch off the local DMD potential and define the fluctuation fields as the expansion around the mean atom-density distribution $\nGP(x)$ in \eqref{Eq:TF} present during the time evolution.
Thus, at $t = 0$, we would find that
$\langle \pif(\cdot, 0) \rangle
= \sqrt{\pi/2} \langle \dd(\cdot, 0)\rangle$
[see \eqref{Eq:Conversion}, or
$\langle \pif(\cdot, 0) \rangle
= \sqrt{\pi} \langle \dd_{\mathrm{r}}(\cdot, 0)\rangle$ for a pair of quasi-condensates]
corresponds to the initial shape of a wave packet located in the region where the atoms have been displaced by the DMD.
Additionally, we can assume that the gas is static (zero mean velocity) and set $\langle \Delta\hat{\phi}(\cdot, 0) \rangle = 0$.
For subsequent times $t > 0$ the wave packet will be propagated through the system according to
\begin{equation}
\langle \pif(x, t) \rangle
= \int_{-R}^{R} \mathrm{d}x'\,
  \left(
    G_{\pi\phi}(x, x'; t) \, \langle \Delta\hat{\phi}(x', 0) \rangle
    + G_{\pi\pi}(x, x'; t) \, \langle \pif(x', 0) \rangle
  \right),
\end{equation}
where due to the initial conditions only
$G_{\pi\pi}(x, x'; t) = G_{\phi\phi}(x, x'; t) + 1/2R$
[see \eqref{G_phi_phi_G_phi_pi} and Appendix~\ref{Appendix:GF}] plays a role.

\begin{figure}[!htbp]

\centering

\includegraphics[trim = 0cm 0cm 0cm 0cm, clip,width=0.9\linewidth]{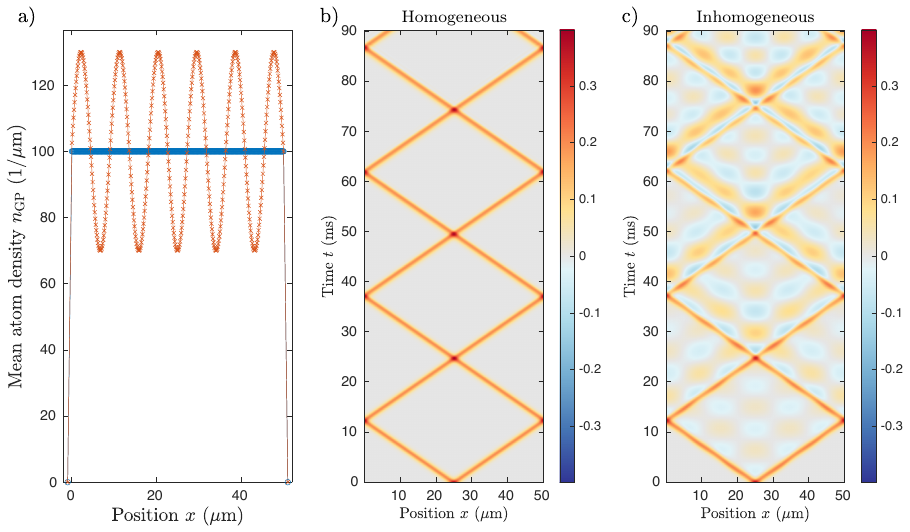}

\caption{%
a) Consider a mean atom distribution corresponding to a homogeneous distribution of 5000 atoms in a box of length $50$ $\mu\mathrm{m}$ (blue circles) and an inhomogeneous one modulated by a sine function with amplitude $30\%$ of the density in the homogeneous case and commensurate with the system size (red crosses).
b) Plot of the spatio-temporal dynamics of the first moment of the density fluctuation $\langle \pif(x, t)\rangle$.
The results of propagating a localized wave packet ($\langle \pif(x, t=0) \rangle$ is a Gaussian function centered in the middle of the system) which is initially static ($\langle \Delta\hat{\phi}(x, t=0) \rangle = 0$). 
The wave packet splits into two counter-propagating packets that repeatedly traverse the system at constant speed.
c) The same initial condition propagated with the inhomogeneous Hamiltonian that arises from the modulated mean-atom distribution.
Again two counter-propagating wave packets are visible but due to the inhomogeneity of the couplings there is no (ballistic) localization anymore.%
The results in both b) and c) are obtained using a lattice discretization approach discussed in \cite{Gluza2020, Gluza2021quantum} with lattice spacing equal to the spacing of the bullets in a).
}%

\label{Fig:Bump_dynamics}

\end{figure}

To give a specific example, let the initial condition for the density fluctuation be
\begin{equation}
\langle \pif(x, 0) \rangle
= C \ee^{-(x - x_0)^2 / 2 \sigma^2_\text{WP}}
\end{equation}
where $C$ is a constant and $\sigma_\text{WP}$ is the range of the action of the DMD potential centered around $x_0$.
This together with $\langle \Delta\hat{\phi}(\cdot, 0) \rangle = 0$ yields
\begin{equation}
\langle \pif(x, t)\rangle
= C \int_{-R}^{R} \mathrm{d}x'\,
  G_{\pi\pi}(x, x'; t) e^{-(x'-x_0)^2 / 2 \sigma^2_\text{WP}}
\end{equation}
for the propagation of the density fluctuation.
After switching off the local DMD potential we expect two wave packets to emanate from the point $x_0$ propagating in opposite directions.
This could then be observed by considering the measurement of the mean atom-density distribution in time and comparing with $\nGP(x)$.
Given the above, we expect that, if $\sigma_\text{WP}$ is smaller than the length scale of the amplitude `leakage' into the light cone due to the Huygens-Fresnel principle violation, then this violation could be observed in a cold-atom experiment; see Fig.~\ref{Fig:Bump_dynamics} for plotted numerical results of the propagation in the presence of a harmonically modulated density distribution.

The example in Fig.~\ref{Fig:Bump_dynamics} uses realistic parameters for state-of-the-art atom-chip experiments \cite{Schweigler_thesis, Tajik2019,Rauer2018}. 
The smallest length-scale in such experiments is the healing length and is in sub-micron regime \cite{Schweigler_thesis}.
In \cite{Tajik2019} a pixel of the DMD effectively changed the light-field potential at $\sigma_\text{DMD} \approx \SI1{\micro\meter}$ such that its effect on the atom density was visible using the absorption imaging system which has a resolution of $\sigma_\text{AIS} \approx \SI{2}{\micro\meter}$.
While $\sigma_\text{AIS}$ is technologically difficult to decrease, $\sigma_\text{DMD}$ could in principle be lowered such that a smoother potential would always be created, but it is practical for these two resolutions to be comparable in order to see the effect of the DMD by the absorption imaging system.
The next length scale $\sigma_\text{WP}$, which corresponds to the size of the wave packets, should be chosen such that one can resolve its shape given $\sigma_\text{AIS}$.
This size can be tuned by programmably selecting enough DMD pixels to create a given shape.
Finally, the largest length scale is the size of the system itself, which for example in \cite{Rauer2018} was $L = \SI{50}{\micro\meter}$.
Fig.~\ref{Fig:Bump_dynamics} shows an example where the density potential used to break the Huygens-Fresnel principle was chosen to be modulated at a relatively small length scale of $\sigma_\text{BHFP} \approx \SI{10}{\micro\meter}$, corresponding to five minima over the length $L = \SI{50}{\micro\meter}$ of the system.
This modulation length scale $\sigma_\text{BHFP}$ is again decided by the DMD resolution $\sigma_\text{DMD}$ on the lower end and by the system size $L$ on the higher end (the parabolic shape being the minimal symmetric example).
In Fig.~\ref{Fig:Bump_dynamics} we have a wave packet with an extension $\sigma_\text{WP}$ that is smaller than the inhomogeneity inducing modulation $\sigma_\text{BHFP}$.
This suffices to illustrate a more intricate spatial pattern signifying the breaking of the Huygens-Fresnel principle than for a parabolic potential.
We point out that the other regime, $\sigma_\text{WP} > \sigma_\text{BHFP}$, can also be considered and is accessible using the DMD technology, but that it may lead to a more complicated interpretation of the results.


\section{Concluding remarks}
\label{Sec:Concluding_remarks}


Motivated by recent atom-chip experiments we studied physical properties of an inhomogeneous generalization of the Tomonaga-Luttinger liquid (TLL) description, referred to as \emph{inhomogeneous TLL}, where the usual propagation velocity and Luttinger parameter are promoted to functions $v(x)$ and $K(x)$ of the spatial coordinate $x$.
While inhomogeneous TLL can be used to effectively describe different experimental setups (see Fig.~\ref{Fig:Illustration}) our main inspiration comes precisely from trapped ultracold atoms on an atom-chip.
Prompted by this application, we restricted our discussion to the family of inhomogeneous TLLs where $v(x)/K(x)$ is constant; this is quite general as it describes the situation where both $v(x)$ and $K(x)$ are given by the mean density of the quantum liquid, which is typically what introduces inhomogeneities in the TLL description.
We showed that this family can be studied by analytical means by expanding the fields in a suitable eigenfunction basis obtained from an associated Sturm-Liouville problem.
As a specific example, we presented explicit results for the special case where this basis is given by Legendre polynomials, which is directly relevant for describing ultracold atoms in a parabolic trap in the weakly interacting regime.

Our emphasis were on non-equilibrium properties, and we showed that inhomogeneous $v(x)$ and $K(x)$ have profound effects on the dynamics.
To identify the cause of each effect, we compared with the case of a wave equation with only an inhomogeneous velocity (and constant Luttinger parameter).
The following three effects were observed:
\begin{enumerate}[label={(\alph*)}, ref={(\alph*)}]

\item
Due to $v(x)$, the usual light cone becomes curved, as observed previously, see, e.g., \cite{Dubail2017lightcones}.

\item
Due to $K(x)$, excitations no longer propagate ballistically localized to the light cone.
Instead, in general, there is spreading inside the cone, meaning that the Huygens-Fresnel principle is not valid.

\item
Due to $K(x)$, there are approximate recurrences, with the same period as the exact recurrences for when only $v(x)$ depends on position.

\end{enumerate}
From our Sturm-Liouville approach, one can observe that the above effects are encoded in the eigenfunctions and the eigenvalues.
For instance, when $K(x)$ depends on position, the eigenfunctions are no longer cosine waves (assuming Neumann boundary conditions), while if only $v(x)$ depends on position, the eigenfunctions can be written as cosine waves by a change of coordinate.
This is also consistent with the path-integral approach where one can observe that $v(x)$ is geometric in nature and can be encoded in the metric describing the emergent curved spacetime, while $K(x)$ cannot and appears in a non-trivial way in the Lagrangian density.

Based on our results, we proposed experimental studies for trapped ultracold atoms on an atom-chip, where we expect that the above effects can be experimentally observed (see Sec.~\ref{Sec:Experimental_observations}).
In addition to the violation of the Huygens-Fresnel principle, one could also investigate the interplay between the inhomogeneities in the TLL description with other effects such as non-linearities appearing at late times due to the underlying interactions, which can be accounted for, e.g., through simulations of the time-dependent Gross-Pitaevskii equation.
Moreover, in an appendix (see Appendix~\ref{App:iTLL_with_mass_term}), we briefly discussed the prospect of using our Sturm-Liouville approach to study inhomogeneous systems with an effective mass term, which can also be realized in atom-chip experiments; in that case, due to the non-trivial coupling of modes, this could allow for selective addressing of specific modes in such systems.

One important question that we did not address concerns how the inhomogeneous TLL phenomenology affects transport, e.g., in nanowires studied in solid-state applications and nanotechnology.
For instance, it would be interesting to study the conductive properties in systems where the Huygens-Fresnel principle is violated due to inhomogeneities that lead to a position-dependent $K(x)$ and to address questions about purely ballistic versus diffusive transport.
This was, e.g., studied in \cite{LangmannMoosavi2019} for the case when only $v(x)$ depends on position through a random function, modeling certain inhomogeneous systems with quenched disorder.
Moreover, it would be interesting to study the effect of such disorder in the general inhomogeneous TLL description presented here.

We note that corresponding results to those obtained here can, in principle, be obtained for a (local) quantum quench in the following sense:
Couple the inhomogeneous TLL theory $H_{\mathrm{iTLL}}$ to an external field producing a given initial wave form, say, in the ground-state expectation of $\Delta\hat{\phi}(x)$ or $\pif(x)$ with respect to the Hamiltonian for the coupled theory, and then turn off the field and let this state evolve under $H_{\mathrm{iTLL}}$.
We did not study such a quench here, but our results in Sec.~\ref{Sec:Dynamics_for_iTLLs:GF} describe the same kind of dynamics.
On this topic, it would be interesting to study a different (global) quantum quench, namely changing $K(x)$ from one function $K_{1}(x)$ to another function $K_{2}(x)$ at time zero, which would be an inhomogeneous generalization of the type of interaction quench studied in \cite{Cazalilla2006, IucciCazalilla2009}.

It would also be interesting to extend our results to finite temperature.
For homogeneous TLLs the combination of finite temperature and system size leads to expressions involving elliptic functions for correlation functions of vertex operators, see \cite{MattssonEtAl1997}.
The zero-mode contribution, since it does not differ between the inhomogeneous setting studied here and homogeneous TLLs (depending on boundary conditions), would again give rise to elliptic functions.
On the other hand, the non-zero-mode contribution is much more complicated in general and would not be expected to feature known special functions, except possibly for certain special cases.

Lastly, in the context of the cold-atom application of TLL, it would be interesting to study the interplay between the effects of a parabolic trap and the cosine interaction resulting from the coupling of two parallel quasi-condensates as outlined in Sec.~\ref{Sec:Atom_chip}.
This can be done by combining the techniques presented here with the Hamiltonian truncation methods \cite{FRT1998, FRT1999, Kukuljan2018correlation, JamesEtAl2018} mentioned in the main text.
Such a study is especially important for understanding some puzzling features of the dynamics in this experimental system.
Specifically, it has been observed that the dynamics exhibits an oscillation damping effect \cite{Phase-locking} with characteristics that are quite unexpected in standard sine-Gordon theory \cite{Horvath2019} which is built as a perturbation of the homogeneous Gaussian massless boson model.
This effect has been recently attributed to the presence of the parabolic trap \cite{NieuwkerkEtAl2021, MennemannEtAl2021}, motivating the above proposed approach.


\begin{acknowledgments}
We thank J{\'e}r\^{o}me Dubail and Paola Ruggiero for useful discussions.
M.G.\ was supported via the European Union's Horizon 2020 research and  innovation programme (PASQuanS, Grant Agreement No.\ 817482).
P.M.\ is grateful for financial support from the Wenner-Gren Foundations (Grant No.\ WGF2019-0061).
S.S.\ was supported by the Slovenian Research Agency (ARRS) under the grant QTE (Grant No.\ N1-0109) and in part by the Foundational Questions Institute (Grant No.\ FQXi-IAF19-03-S2).
\end{acknowledgments}


\begin{appendix}


\section{Lagrangian density in curved spacetime}
\label{App:Lagrangian_in_curved_spacetime}


In this appendix we derive the general Lagrangian density $\mathcal{L}$ in curved spacetime that corresponds to the Hamiltonian
\begin{equation}
\label{Eq:H_iTLL_massive}
H
= \frac{1}{2} \int_{-R}^{R} \mathrm{d}x\,
    \biggl(
	    \frac{v(x)}{K(x)} \pif(x)^2
		+ v(x)K(x) [\partial_{x} \hat{\phi}(x)]^2
		+ (M\v)^2 v(x)K(x) \hat{\phi}(x)^2
	\biggr)
\end{equation}
for general $v(x)$ and $K(x)$.
(For simplicity, we omit normal ordering in this appendix.)
When $M = 0$, this is the same as $H$ in \eqref{Eq:H_iTLL}, in which case we show that $\mathcal{L}$ is given by \eqref{cL_iTLL} with the associated metric.

Let $(h_{\mu\nu})$ be an unknown metric tensor with Lorentzian signature $(+,-)$ in the coordinates $(x^0, x^1) = (\v t, x)$ for some arbitrary velocity parameter $\v$, i.e., $\mathrm{d}s^2 = h_{\mu\nu} \mathrm{d}x^{\mu} \mathrm{d}x^{\nu}$.
In the path-integral formulation, we postulate the action $S = \int_{-\infty}^{\infty} \mathrm{d}t \int_{-R}^{R} \mathrm{d}x\, \mathcal{L}$ with
\begin{equation}
\mathcal{L}
= \frac{\v}{2} \sqrt{-h}
  \Bigl[
    A(x) h^{\mu\nu} (\partial_{\mu} \phi) (\partial_{\nu} \phi)
	- (M\v)^2 B(x) \phi^2
  \Bigr],
\end{equation}
where $A(x)$ and $B(x)$ are two unknown functions of $x$.
We recall that
$\partial_{\mu} = \partial/\partial x^{\mu}$, 
$h = \det( h_{\mu\nu} ) < 0$,
and
$h^{\mu\nu} h_{\nu\sigma} = \delta^{\mu}_{\sigma}$.
Let $\mathcal{E} = \mathcal{E}(x)$ given by
\begin{equation}
\mathcal{E}(x)
= \frac{1}{2}
  \biggl(
    \frac{v(x)}{K(x)} \pi_{\phi}(x)^2
	+ v(x)K(x) [\partial_{x} \phi(x)]^2
	+ (M\v)^2 v(x) K(x) \phi(x)^2
  \biggr)
\end{equation}
be the Hamiltonian density corresponding to $H$ in \eqref{Eq:H_iTLL_massive}.
(For clarity, we write $\pi_{\phi}$ with a subscript to indicate that it is conjugate to the field $\phi$.)
Then
\begin{equation}
\label{cL_from_cE}
\mathcal{L}
= \v \pi_{\phi} \partial_0 \phi - \mathcal{E},
\qquad
\pi_{\phi}
= \frac{1}{\v} \frac{\partial \mathcal{L}}{\partial (\partial_{0}\phi)}.
\end{equation}
On general grounds,
\begin{equation}
\label{pi_phi_C(x)}
\pi_{\phi}
= C(x) \partial_{0} \phi
\end{equation}
for some unknown function $C(x)$.
We thus want to determine $A(x)$, $B(x)$, $C(x)$, and $h_{\mu\nu}$ such that \eqref{cL_from_cE} is satisfied.
Since the metric must be diagonal for this to be true, the conditions in \eqref{cL_from_cE} are equivalent to the system of equations:
\begin{subequations}
\label{eq_sys}
\begin{align}
A(x) \sqrt{ \frac{-h_{11}}{h_{00}} }
& = C(x) \biggl[ 2 - \frac{v(x)}{\v K(x)} C(x) \biggr],
& \sqrt{-h_{00} h_{11}} B(x)
& = \frac{v(x)K(x)}{v}, \\
A(x) \sqrt{ \frac{h_{00}}{-h_{11}} } 
& = \frac{v(x)K(x)}{\v},
& 2 - \frac{v(x)}{\v K(x)} C(x)
& = 1.
\end{align}
\end{subequations}
The possible solutions are
\begin{equation}
A(x) = K(x),
\qquad
B(x) = \frac{K(x)}{-h_{11}},
\qquad
C(x) = \frac{\v K(x)}{v(x)},
\qquad
h_{00}
= \biggl( \frac{v(x)}{\v} \biggr)^2 (-h_{11}),
\end{equation}
where the only freedom is between $B(x)$, $h_{00}$, and $h_{11}$.
Since all two-dimensional Lorentzian manifolds are conformally equivalent, we can without loss of generality set $h_{11} = -1$, which implies
\begin{equation}
\label{cL_iTLL_massive}
\mathcal{L}
= \frac{\v}{2} \sqrt{-h} K(x) 
  \Bigl[
    h^{\mu\nu} (\partial_{\mu} \phi) (\partial_{\nu} \phi)
	- (M\v)^2 \phi^2
  \Bigr],
\qquad
(h_{\mu \nu})
= \left( \begin{matrix}
    v(x)^2/\v^2 & 0 \\
    0          & -1 
	\end{matrix} \right).
\end{equation}
In particular, the above gives \eqref{cL_iTLL} when $M = 0$.
Note that the parameter $\v$ above is arbitrary for general $v(x)$ and $K(x)$, while it can be chosen to be given by \eqref{vK_condition} if that condition is satisfied.

As a closing remark, note that \eqref{cL_iTLL} agrees with Eq.~(10) in \cite{BrunDubail2018igff} since interchanging $\partial_{x} \hat{\phi}(x)$ and $\pif(x)$ corresponds to replacing $K(x)$ by $K(x)^{-1}$.
For the homogeneous case, but which generalizes to inhomogeneous iTLLs, one way to observe this duality is to compare \eqref{Eq:H_iTLL} with Eq.~(3.25) in \cite{Giamarchi2003}, or explicitly by comparison with the corresponding formulas in \cite{LangmannMoosavi2015} where the notation is such that
$\partial_{x} \Phi(x) = (1/\sqrt{\v}) \pif(x)$
and
$\Pi(x) = \sqrt{\v} \partial_{x} \hat{\phi}(x)$.


\section{Diagonalization of the inhomogeneous TLL Hamiltonian with a mass term}
\label{App:iTLL_with_mass_term}


In this appendix we discuss the diagonalization of the Hamiltonian [cf.\ \eqref{Eq:H_sG} and the argument leading to \eqref{Eq:H_iTLL_vK_ibp}]
\begin{equation}
\label{H_F_M}
H_{F, M}
= \frac{\v}{2} \int_{-R}^{R} \mathrm{d}x\,
  \Bigl(
    \pif(x)^{2}
    - \hat{\phi}(x) \partial_{x} [F(x) \partial_{x} \hat{\phi}(x)]
    + (M\v)^2 F(x) \hat{\phi}(x)^{2}
  \Bigr),
\end{equation}
which contains an inhomogeneous mass term if $M \neq 0$.
(For simplicity, we omit normal ordering in this appendix.)
This Hamiltonian can be diagonalized as before using Sturm-Liouville theory for the new operator
\begin{equation}
\mathcal{A}'
= - \partial_{x} F(x) \partial_{x} + (M\v)^2 F(x)
\label{SL_op_Ap}
\end{equation}
with suitable boundary conditions.
(Note that if the Sturm-Liouville problem is regular for $M = 0$, then it is also regular for $M \neq 0$.)

Even though Sturm-Liouville theory guarantees that there is a complete basis of orthonormal eigenfunctions, it may be non-trivial to construct them and find the corresponding eigenvalues for the more general case above.
For this reason it is useful to (formally) solve this problem by expressing the eigenmodes of the Hamiltonian for $M \neq 0$ as linear combinations of those for $M = 0$.
(This approach is formal since we do not address questions about unitary equivalence between the Hilbert spaces for $M = 0$ and $M \neq 0$ nor do we discuss how this affects the zero modes, since this is beyond the scope of this paper.)

We expand the Hamiltonian \eqref{H_F_M} using the eigenfunctions $f_{n}(x)$ corresponding to $H_{F} = H_{F, M = 0}$, which yields
\begin{equation}
H_{F, M}
= H_{F, M = 0}
  + \frac{M^2 \v^3}{2}
    \sum_{n, n' = 0}^{\infty}
    J_{n, n'} \hat{\phi}_{n} \hat{\phi}_{n'},
\end{equation}
where we defined
\begin{equation}
J_{n, n'}
= \int_{-R}^{R}\mathrm{d}x\,  F(x) f_{n}(x) f_{n'}(x).
\label{couplings}
\end{equation}
By expanding $F(x)$ in these eigenfunctions, $F(x) = \sum_{n = 0}^{\infty} f_{n}(x) F_{n}$, and inserting this into \eqref{couplings}, we obtain
\begin{equation}
J_{n, n'}
= \sum_{n'' = 0}^{\infty} F_{n''}
  \int_{-R}^{R}\mathrm{d}x\, f_{n}(x) f_{n'}(x) f_{n''}(x).
\end{equation}
From this we can read off a number of properties.
First of all, if $F_{n''}$ decays sufficiently slow, then $J_{n, n'}$ is small for $n, n' \ll n''$, since the integral is oscillatory.
Similarly, if $F_{n''}$ decays sufficiently fast, then $J_{n, n'}$ is small for $n, n' \gg n''$, again since the integral is oscillatory.
This property is important and shows that the effect of adding the mass term affects most strongly the low-lying modes.

From the above, we see that the eigenmodes for $H_{F, M \neq 0}$ can either be found by directly solving the associated Sturm-Liouville problem or by attempting to diagonalize the infinite matrix obtained from the Hamiltonian expressed in terms of the eigenfunctions $f_{n}(x)$; the latter is in general non-diagonal due to $J_{n, n'}$. 
Numerical solutions of the Sturm-Liouville problem essentially rely on the second approach of expanding the problem in convenient basis functions, truncating, and solving the resulting matrix problem.

As a first simple example, one can consider the case of a constant $F(x) = F$ with Neumann boundary conditions.
In this case, the eigenfunctions for $M \neq 0$ are given by cosine functions, $F_{n} \propto \delta_{n, 0}$, and $J_{n, n'} \propto \delta_{n, n'}$ (the latter follows from that the eigenfunctions are orthonormal).

As a second example, we consider our special case in \eqref{K_parabolic}.
Then $F(x) = 1 - (x/R)^{2}$, for which we recall that the energies $E_{n}$ and the eigenfunctions $f_{n}(x)$ are given by \eqref{E_n} and \eqref{iTLL_parabolic_lambda_n_f_n} obtained as solutions to Legendre's differential equation.
From \eqref{couplings} it follows that
\begin{equation}
J_{n, n'}
= \sqrt{\frac{(2n+1)(2n'+1)}{4}}  
  \int_{-1}^{1} \mathrm{d}\zeta\, (1 - \zeta^2) P_{n}(\zeta) P_{n'}(\zeta)
\end{equation}
(after a change of variables to $\zeta = x/R$).
Using that $1 - \zeta^2 = (2/3)[P_{0}(\zeta) - P_{2}(\zeta)]$, this can be computed with the help of a so-called Wigner 3-$j$ symbol \cite[\href{http://dlmf.nist.gov/34.3}{{\S}34.3}]{NIST:DLMF} to express the integral of a product of three Legendre polynomials:
\begin{subequations}
\begin{gather}
\int_{-1}^{1} \mathrm{d}\zeta\, (1 - \zeta^2) P_{n}(\zeta) P_{n'}(\zeta)
= c_{n} \delta_{n, n'}
  + d_{n} \delta_{n, n'-2}
  + d_{n'} \delta_{n', n-2}, \\
c_{n}
= \frac{4(n^{2}+n-1)}{(2n+1)(2n-1)(2n+3)},
\qquad
d_{n}
= - \frac{2(n+1)(n+2)}{(2n+1)(2n+3)(2n+5)}.
\end{gather}
\end{subequations}
It follows that
\begin{equation}
\int_{-R}^{R} \mathrm{d}x\, F(x) \hat{\phi}(x)^{2}
= \sum_{n = 1}^{\infty}
  \biggl(
    \frac{2n+1}{2} c_{n}
    \hat{\phi}_{n}^{2}
    + 2 \sqrt{\frac{(2n+1)(2n+5)}{4}} d_{n}
      \hat{\phi}_{n} \hat{\phi}_{n+2}
  \biggr)
  + (\text{zero modes})
\end{equation}
which inserted into \eqref{Eq:H_iTLL_vK_ibp} together with \eqref{H_F_diagonalized} yields
\begin{align}
\label{H_F_M_matrix_rep}
H_{F, M}
= \sum_{n, n' = 1}^{\infty}
  \Bigl(
    \xi_{n, n'}
    \Bigl[
      \hat{a}_{n}^{\dagger} \hat{a}_{n'}\pdag
      + \hat{a}_{n}\pdag \hat{a}_{n'}^{\dagger}
    \Bigr]
    + \eta_{n, n'}
      \Bigl[
        \hat{a}_{n}^{\dagger} \hat{a}_{n'}^{\dagger}
        + \hat{a}_{n}\pdag \hat{a}_{n'}\pdag
      \Bigr]
    \Bigr)
  + (\text{zero modes})
\end{align}
with
\begin{subequations}
\begin{gather}
\xi_{n, n'}
= \frac{1}{2} E_{n} \delta_{n, n'}
  + \frac{M^2 \v^3}{4} K_{n, n'},
\qquad
\eta_{n, n'}
= \frac{M^2 \v^3}{4} K_{n, n'}, \\
K_{n, n'}
= \frac{1}{2E_{n}} \frac{2n+1}{2} c_{n} \delta_{n, n'}
  + \frac{1}{2}
    \left[
      \frac{\sqrt{(2n+1)(2n+5)}}{2\sqrt{E_{n}E_{n+2}}}
      d_{n} \delta_{n, n'-2}
      + (n \leftrightarrow n')
    \right]
\end{gather}
\end{subequations}
for $n, n' = 1, 2, \ldots$.
(We recall that the zero modes require a separate treatment and are thus not explicitly considered.)

The Hamiltonian in \eqref{H_F_M_matrix_rep} can be diagonalized by means of a multi-mode (in reality, an infinite-mode) Bogoliubov transformation.
Note that the infinite matrix $(K_{n, n'})$ is symmetric and 5-diagonal with non-constant coefficients.
More precisely, $(K_{n, n'})$ is non-zero only along three parallel diagonals starting from the elements at $(n,n') = (1,1)$, $(1,3)$, and $(3,1)$.
Moreover, its elements decay for large $n$ since $c_{n} \sim d_{n} \sim 1/n$.
This means that if we truncate the infinite matrix up to a maximum dimension $N \times N$, diagonalize the resulting finite matrix, and repeat for increasing values of $N$, the spectra obtained in this way can be  expected to converge for large $N$ to the exact spectrum of $H_{F, M}$.

\begin{figure}[!htbp]

\centering

\subfigure[]{%
\includegraphics[width=0.4\linewidth]{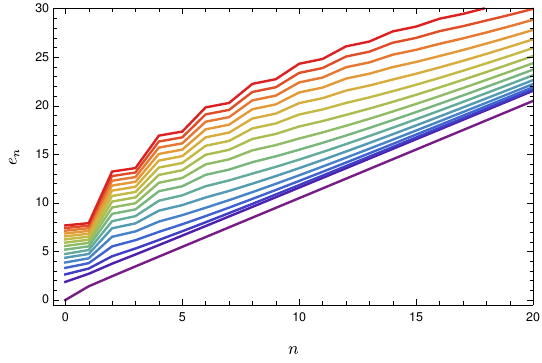}
\label{Fig:TF_spectra:plotJspectra}
}%
\hspace{10mm}
\subfigure[]{%
\includegraphics[width=0.4\linewidth]{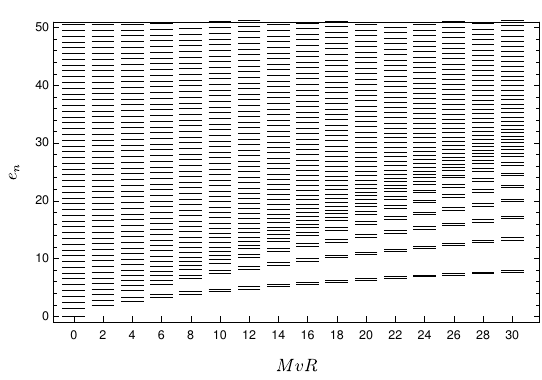}
\label{Fig:TF_spectra:plotJspectra2}
}%

\caption{%
\subref{Fig:TF_spectra:plotJspectra}
Lowest energy eigenvalues $e_{n} = E_{n} R/v$ for $H_{F, M}$ for $M\v R = 0, 2, 4, \ldots, 30$ (bottom to top curve) computed numerically (as described in the text) truncating after $N = 1000$ modes.
\subref{Fig:TF_spectra:plotJspectra2} The same results illustrated differently.%
}%

\label{Fig:TF_spectra}

\end{figure}
\begin{figure}[!htbp]

\centering

\subfigure[Ground-state energy]{%
\includegraphics[width=0.4\linewidth]{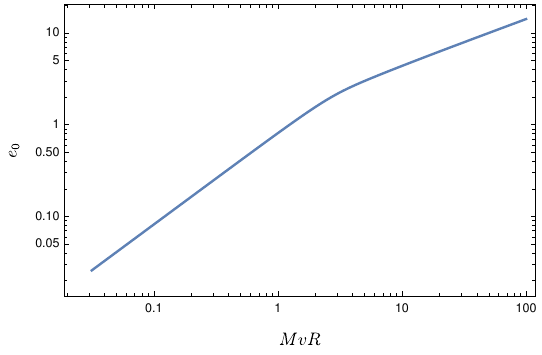}
\label{Fig:TF_spectra2:Left}
\hspace{10mm}
}%
\subfigure[Energy gap]{%
\includegraphics[width=0.4\linewidth]{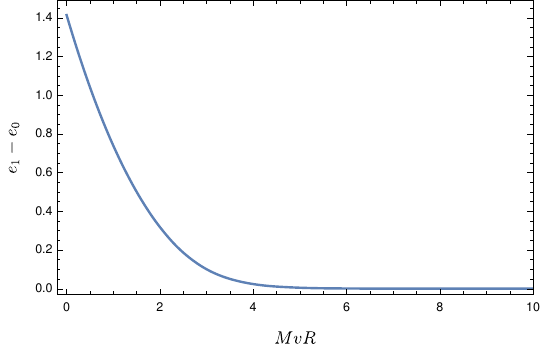}
\label{Fig:TF_spectra2:Right}
}%

\caption{%
\subref{Fig:TF_spectra2:Left} Log-log plot of the ground-state energy $e_{0} = E_{0}R/v$ and \subref{Fig:TF_spectra2:Right} plot of the energy gap $e_{1} - e_{0} = (E_{1} - E_{0})R/v$ for $H_{F, M}$ as functions of $M\v R$ computed numerically (as described in the text) truncating after $N = 1000$ modes.%
}%

\label{Fig:TF_spectra2}

\end{figure}
\begin{figure}[!htbp]

\centering

\includegraphics[width=0.6\linewidth]{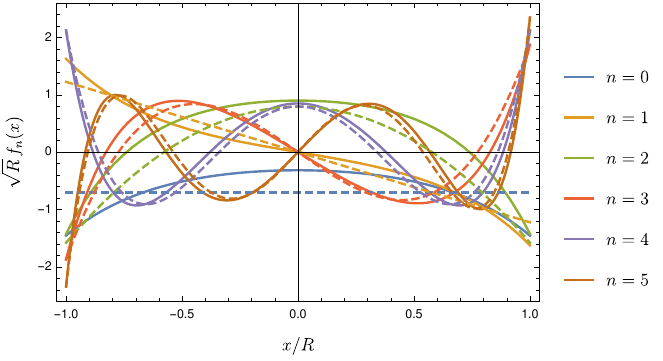} 

\caption{%
Lowest eigenfunctions $f_{n}(x)$ for $H_{F, M}$ for $M\v R = 3$ (solid lines) and $M\v R = 0$ (dashed lines) computed numerically (as described in the text) truncating after $N = 1000$ modes.%
}%

\label{Fig:TF_spectra3}

\end{figure}

Fig.~\ref{Fig:TF_spectra} shows the lowest energy eigenvalues of $H_{F, M}$ computed numerically as described above for different values of $M\v R$. 
For sufficiently large $M\v R$, the lowest levels come in approximate pairs of nearby values, while the highest levels are closer and closer to those for $H_{F, M = 0}$.
The former can be explained by noticing that, for large $M\v R$, the derivative term of the Sturm-Liouville operator is less significant compared to the mass term, which forces the eigenfunctions to become more vanishing everywhere except near the boundaries.
Thus, the eigenfunctions differ mainly by contributions near the boundaries, with spatially symmetric or anti-symmetric eigenfunctions having approximately the same energy, giving rise to the observed approximate double degeneracy of the levels.
Fig.~\ref{Fig:TF_spectra2} shows the dependence of the ground-state energy and the energy gap (between the first excited state and the ground state) on $M\v R$.
The ground state energy appears to follow two different types of power-law behavior:
$E_0 \sim M\v R$ for $M\v R < 1$ and $E_0 \sim \sqrt{M\v R}$ for $M\v R > 1$ (based on curve fitting).
Lastly, Fig.~\ref{Fig:TF_spectra3} shows a comparison of the lowest eigenfunctions $f_{n}(x)$ for $M\v R = 3$ with those for $M\v R = 0$.
The relative difference is significant for small $n$ but decreases for larger $n$.


\section{Formulas for Green's functions}
\label{Appendix:GF}


In this appendix we give details for the Green's functions for
$\pif(x, t)$
and
$\partial\hat{\phi}(x, t) = \partial_{x} \hat{\phi}(x, t)$.

If follows from \eqref{dt_phi_dt_pi} that the Heisenberg equation of motion for the $\pif$ field is of the same form as \eqref{EoM_phi}:
\begin{equation}
\label{EoM_pi}
\partial_{t}^{2} \pif
= \v^{2} \partial_{x} \left[ F(x) \partial_{x} \pif \right].
\end{equation}
Repeating the steps in Sec.~\ref{Sec:Dynamics_for_iTLLs:GF}, we obtain
\begin{equation}
\pif(x, t)
= \int_{-R}^{R} \mathrm{d}x'\,
  \left(
    G_{\pi\pi}(x, x'; t)\, \pif(x', 0)
    + G_{\pi\phi}(x, x'; t)\, \Delta\hat{\phi}(x', 0)
  \right),
\end{equation}
where
\begin{equation}
\label{G_pi_pi_G_pi_phi}
G_{\pi\pi}(x, x'; t)
= G_{\phi\phi}(x, x'; t) + \frac{1}{2R},
\qquad
G_{\pi\phi}(x, x'; t)
= - \frac{1}{\v} \sum_{n = 1}^{\infty} E_{n}^2 f_{n}(x) f_{n}(x') g_{n}(t),
\end{equation}
using $G_{\phi\phi}(x, x'; t)$ in \eqref{G_phi_phi_G_phi_pi} and $g_n(t)$ in \eqref{g_n_t}.
[In the above we used that $f_{0}(x) = 1/\sqrt{2R}$ and $E_{0} = 0$.]

Acting with $\partial_{x}$ on both sides of \eqref{EoM_phi}, the Heisenberg equation of motion for the $\partial\hat{\phi}$ field is
\begin{equation}
\partial_{t}^{2} \partial\hat{\phi}
= \v^{2} \partial_{x}^2 \bigl[ F(x) \partial\hat{\phi} \bigr].
\label{EoM_dphi}
\end{equation}
The general solution in position space is given by \eqref{dphi_GG} together with $G_{\partial\phi\partial\phi}(x, x'; t)$ and $G_{\partial\phi\pi}(x, x'; t)$ in \eqref{G_dphi_dphi_G_dphi_pi}.
[The computation includes a boundary term $+\int_{-R}^{R} \mathrm{d}x'\, \partial_{x} G_{\phi\phi}(x, x'; t) \hat{\phi}(R, 0)$ that can be shown to be zero:
Each term in the sum over $n > 0$ in this boundary term, obtained by inserting $G_{\phi\phi}(x, x'; t)$ in \eqref{G_phi_phi_G_phi_pi} and changing order of the integral and the sum, is proportional to $\int_{-R}^{R} \mathrm{d}x'\, f_{n}(x') = 0$, which follows from \eqref{ON_basis} since $f_{0}(x)$ is constant.]
The Green's functions can explicitly be written as
\begin{equation}
G_{\partial\phi\partial\phi}(x, x'; t)
= - \sum_{n = 1}^{\infty}
    f^{(1)}_{n}(x) f^{(-1)}_{n}(x') \partial_{t} g_{n}(t),
\qquad
G_{\partial\phi\pi}(x, x'; t)
= \v \sum_{n = 1}^{\infty}
  f^{(1)}_{n}(x) f_{n}(x') g_{n}(t),
\end{equation}
where
$f^{(1)}_{n}(x) = \partial_{x} f_{n}(x)$
and
$f^{(-1)}_{n}(x) = \int_{-R}^x \mathrm{d}x'\, f_{n}(x')$.

Using analogous arguments, one can also show that 
\begin{equation}
G_{\pi\partial\phi}(x, x'; t)
= - \int_{-R}^{x'} \mathrm{d}x''\, G_{\pi\phi}(x, x''; t)
= \frac{1}{v} \sum_{n = 1}^{\infty}
    E_{n}^2 f_{n}(x) f^{(-1)}_{n}(x') g_{n}(t).
\end{equation}

In the special case $F(x) = 1 - (x/R)^2$, the eigenfunctions $f_{n}(x)$ are given in \eqref{iTLL_parabolic_lambda_n_f_n}, and, using standard formulas for Legendre polynomials,
\begin{subequations}
\begin{align}
f^{(1)}_{n}(x)
& = (n+1)
    \sqrt{\frac{2n+1}{2R}}
    \frac{x P_n(x/R)-R P_{n+1}(x/R)}{R^2-x^2}, \\
f^{(-1)}_{n}(x)
& = R \sqrt{\frac{2n+1}{2R}}
    \frac{P_{n+1}(x/R) - P_{n-1}(x/R)}{2n+1},
\end{align}
\end{subequations}
where we set $P_{-1}(\cdot) = 0$ (so that these formulas are valid for all $n = 0, 1, \ldots$).


\end{appendix}


%


\end{document}